\documentclass{JHEP3}
\usepackage{graphicx}
\usepackage{epsfig,amsmath,amsthm,axodraw4j}
\newcommand{\be}{\begin{equation}}
\newcommand{\ee}{\end{equation}}
\newcommand{\bea}{\begin{eqnarray}}
\newcommand{\eea}{\end{eqnarray}}
\def\bse{\begin{subequations}}
\def\ese{\end{subequations}}

\def\IZ{\relax\ifmmode\hbox{Z\kern-.4em Z}\else{Z\kern-.4em Z}\fi}

\newcommand{\non}{\nonumber \\}

\def\half{\frac{1}{2}} 

\def\del{{\partial}}

 \def\co{{\cal O}}

\def\al{\alpha} \def\bt{\beta}
 \def\dl{\delta}

\def\presub{\vspace{.5cm} \noindent}

\def\bi{\begin{itemize}} \def\ei{\end{itemize}}

\def\Schw{Schwarzschild }
\def\({\left(} \def\){\right)}
\def\[{\left[} \def\]{\right]}

\title{ \center{Dressing the Post-Newtonian two-body problem \\
 and Classical Effective Field Theory}}

\author{Barak Kol and Michael Smolkin\\
Racah Institute of Physics, Hebrew University\\
Jerusalem 91904, Israel\\
E-mail:
{\tt\href{mailto:barak_kol@phys.huji.ac.il}{barak\_kol@phys.huji.ac.il}}
,\email{smolkinm@phys.huji.ac.il}}

\abstract{We apply a dressed perturbation theory to better
organize and economize the computation of high orders of the
2-body effective action of an inspiralling Post-Newtonian
gravitating binary. We use the effective field theory approach
with the non-relativistic field decomposition (NRG fields). For
that purpose we develop quite generally the dressing theory of a
non-linear classical field theory coupled to point-like sources.
We introduce dressed charges and propagators, but unlike the quantum theory there are no dressed bulk vertices.
    The dressed quantities are found to obey recursive integral equations which succinctly encode parts
of the diagrammatic expansion, and are the classical version of the Schwinger-Dyson equations. Actually,
the classical equations are somewhat stronger since they involve
only finitely many quantities, unlike the quantum theory.
    Classical diagrams are shown to factorize exactly when they
contain non-linear world-line vertices, and we classify all the
possible topologies of irreducible diagrams for low loop numbers.
    We apply the dressing program to our Post-Newtonian case of interest. The dressed
charges consist of the dressed energy-momentum tensor after a
non-relativistic decomposition, and we compute all dressed charges
(in the harmonic gauge) appearing up to 2PN in the 2-body
effective action (and more). We determine the irreducible skeleton
diagrams up to 3PN and we employ the dressed charges to compute
several terms beyond 2PN.}

\begin{document}

\section{Introduction and Summary}

Gravitational wave observatories (see for example the reviews
\cite{IFO-rev} and references therein) demand knowledge of the
waveform emitted by an inspiralling binary system of compact
objects. Fully Generally Relativistic numerical simulations can
now simulate such waveforms - see the review \cite{PretoriusRev}
and references therein. Yet, as always, an analytic treatment is
complementary and improves insight, especially into the functional
dependence of the results on the parameters. A \emph{
perturbative} analytic treatment is possible in two limits. The
first is the Post-Newtonian (PN) approximation, and it holds
whenever the velocities are small compared with the speed of
light, or equivalently through the virial theorem whenever the
separation between the compact objects is much larger than their
\Schw radii. The second limit is that of an extreme mass ratio. In
this paper we shall concentrate on the PN approximation which is
always valid at the initial stages of any inspiral.

The computation of the effective 2-body action in PN was a subject
of considerable research over the last decades and the current
state of the art is its determination up to order 3.5PN, as
summarized in the review \cite{BlanchetRev} (see also the recent
\cite{Schaefer-account}).
Another approach, the effective field theory (EFT) approach to
this problem was suggested by Goldberger and Rothstein (2004)
\cite{GoldbergerRothstein1}, where more traditional GR methods are
replaced by field theoretic tools including Feynman diagrams,\footnote{See also \cite{Okamura:1973my} for certain early days efforts to calculate classical potentials to higher post-Newtonian
orders using field theoretical methods and Feynman graph techniques.} 
loops and regularization. In particular
\cite{GoldbergerRothstein1} reproduced the 1PN effective action
(known as Einstein-Infeld-Hoffmann) within the EFT approach. In
\cite{CLEFT-caged,NRG} the metric components in the
Post-Newtonian, non-relativistic limit  were conveniently
decomposed into a scalar Newtonian potential, a gravito-magnetic
3-vector potential and a symmetric 3-tensor. These fields were
termed collectively ``fields of Non-Relativistic Gravitation (NRG
fields)'', and the derivation of 1PN was shown to further
simplify. In \cite{GilmoreRoss} the 2PN expression was reproduced
within the effective field theory approach together with NRG
fields.

Other recent developments related to either PN or the EFT approach
include: dissipative effects and effective horizon degrees of
freedom \cite{GoldbergerRothstein2,dof}; thermodynamics of caged
black holes \cite{H4,dialogue1} through the EFT approach
\cite{CGR,CLEFT-caged,GRS-caged}; EFT \cite{PortoRothstein06} (see also \cite{Porto:2005ac,Levi:2008nh}) 
Hamiltonians \cite{SC,BRB} for rotating point-particles;
 tidal effects for compact objects \cite{Damour-etal-tidal,BinningtonPoisson};
approximate solutions for higher
dimensional black object including rings \cite{High-d-rings} and
blackfolds \cite{Blackfolds};  a mechanized EFT computation for
2PN \cite{Chu2PN}; ``De-Turek'' gauge for numerical relativity
\cite{HeadrickKitchenWiseman}; and finally radiation reaction and waves within
the EFT approach \cite{CardosoDiasFig,GalleyTiglio}.

\begin{figure}[t!]
\centering \noindent
\includegraphics[width=2cm]{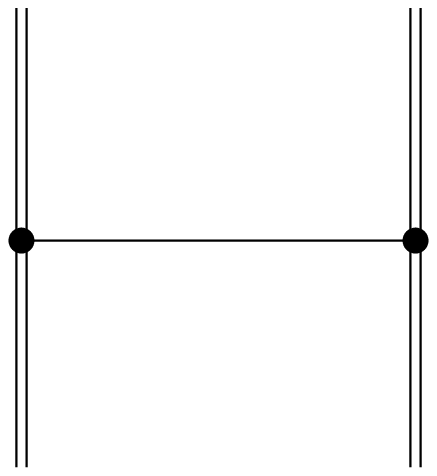}
\caption[]{The diagram which represents the Newtonian potential
interaction mediated through the Newtonian scalar field $\phi$.
The notations will be fully defined later in section
\ref{dressing-PN-section}.}
 \label{0PN-fig}
\end{figure}

\begin{figure}[t!]
\centering \noindent
\includegraphics[width=3cm]{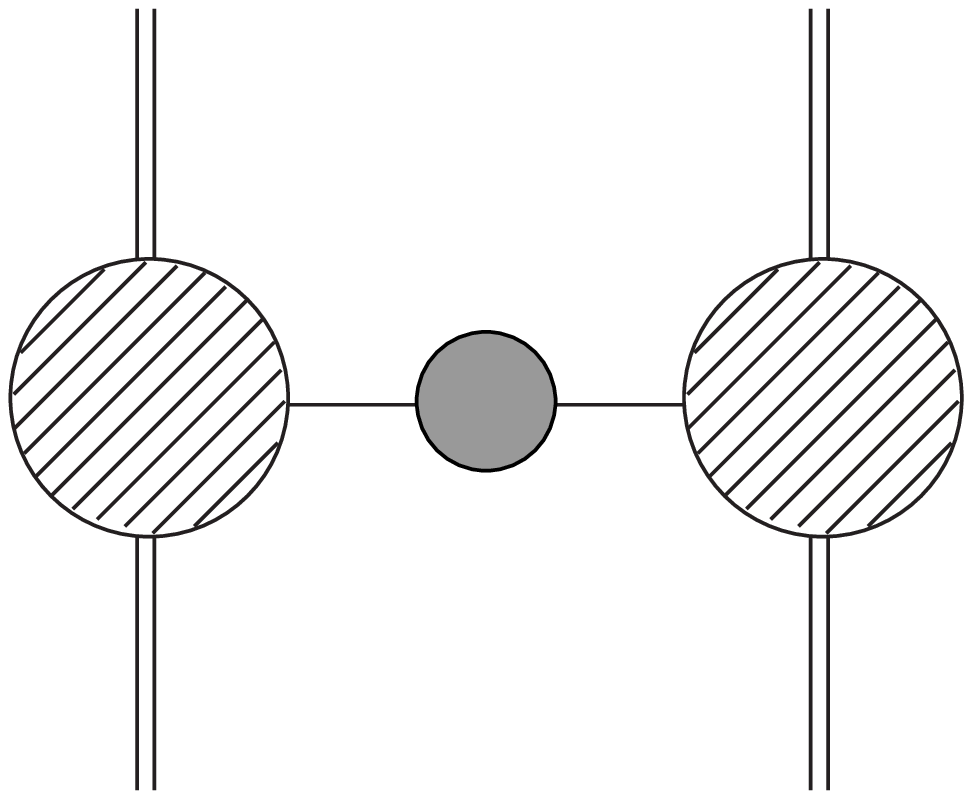}
\caption[]{A class of diagrams that we interpret as the Newtonian
potential interaction between \emph{dressed} energy distributions
of each body through a dressed propagator. The dark blobs
represent any sub-diagram with an arbitrary number of vertices on
the world-line and a single external leg for the Newtonian
potential. There are no bulk loops, as always in classical
physics. The light blob represents any sub-diagram with two
external legs of the Newtonian potential, which amounts to a
propagator with an arbitrary number of retardation insertions.}
\label{0PN-renorm-fig}
\end{figure}

\begin{figure}[t!]
\centering \noindent
\includegraphics[width=8cm]{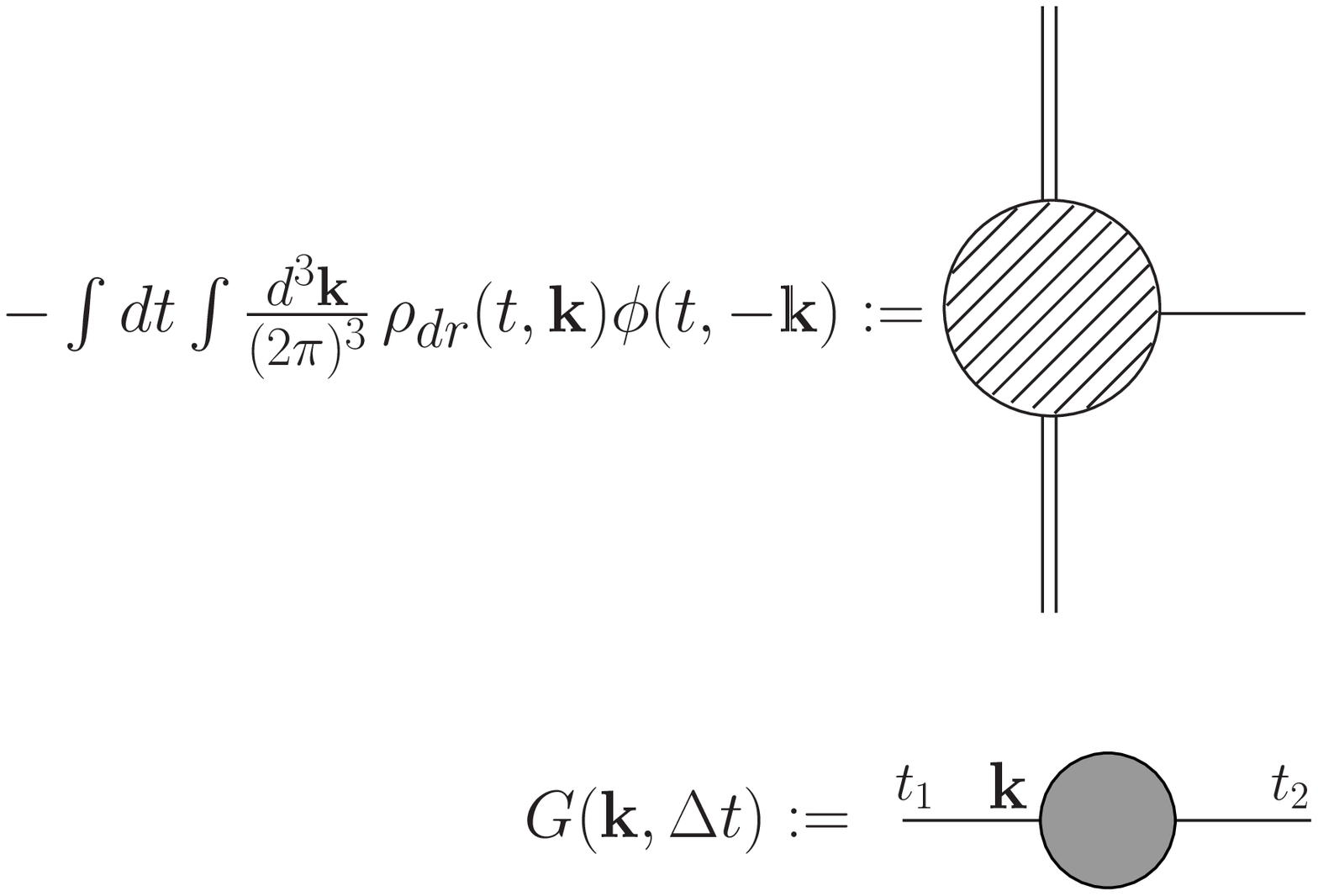}
\caption[]{The diagrammatic definition of the dressed energy
distribution (top) and the dressed propagator (bottom). The
dressed energy distribution is defined through the one point
function for $\phi$ in the presence of a single source (after
stripping external propagators), while the dressed propagator is
defined through the full two point function for $\phi$.}
\label{intro-renorm-fig}
\end{figure}

The computation of order 2PN \cite{GilmoreRoss} demonstrates a
proliferation in the number of diagrams: from 4 at 1PN to 21 at
2PN, and furthermore the number is expected to continue and grow
at the next order. Clearly, it would be useful to have an improved
perturbative expansion. In this paper we shall present a new
method to better organize the calculation and economize it.
 The basic idea is to recognize recurring sub-diagrams which
physically describe ``dressed'' charges and propagators.
The simplest example is furnished by the Newtonian potential $G\,
m_1\, m_2/|r_1-r_2| \subset S_{eff}$
 which belongs to order 0PN. This
term is associated with the diagram in figure \ref{0PN-fig}. At
higher orders there is a class of diagrams of the form shown in
figure \ref{0PN-renorm-fig} where the point masses are replaced by
dressed energy distributions defined on the top row of figure
\ref{intro-renorm-fig} \be
 m_i\, \delta(x_i-r_i) \to \rho_{dr}(x_i-r_i) \ee
and the bare propagator is replaced by the dressed (relativistic) one defined on the bottom of figure \ref{intro-renorm-fig}  \be
 \frac{1}{k^2}~ \delta(\Delta t) \to G(k,\Delta t) ~. \ee
 The same sub-diagrams which define the dressed energy distribution
and appear in the calculation of the dressed Newtonian interaction
(figure \ref{0PN-renorm-fig}) appear also in other diagrams. It
makes sense to record the values of these sub-diagrams and later
re-use them. This is the basic idea of the dressed perturbation
theory.

\begin{figure}[t!]
\centering \noindent
\includegraphics[width=6cm]{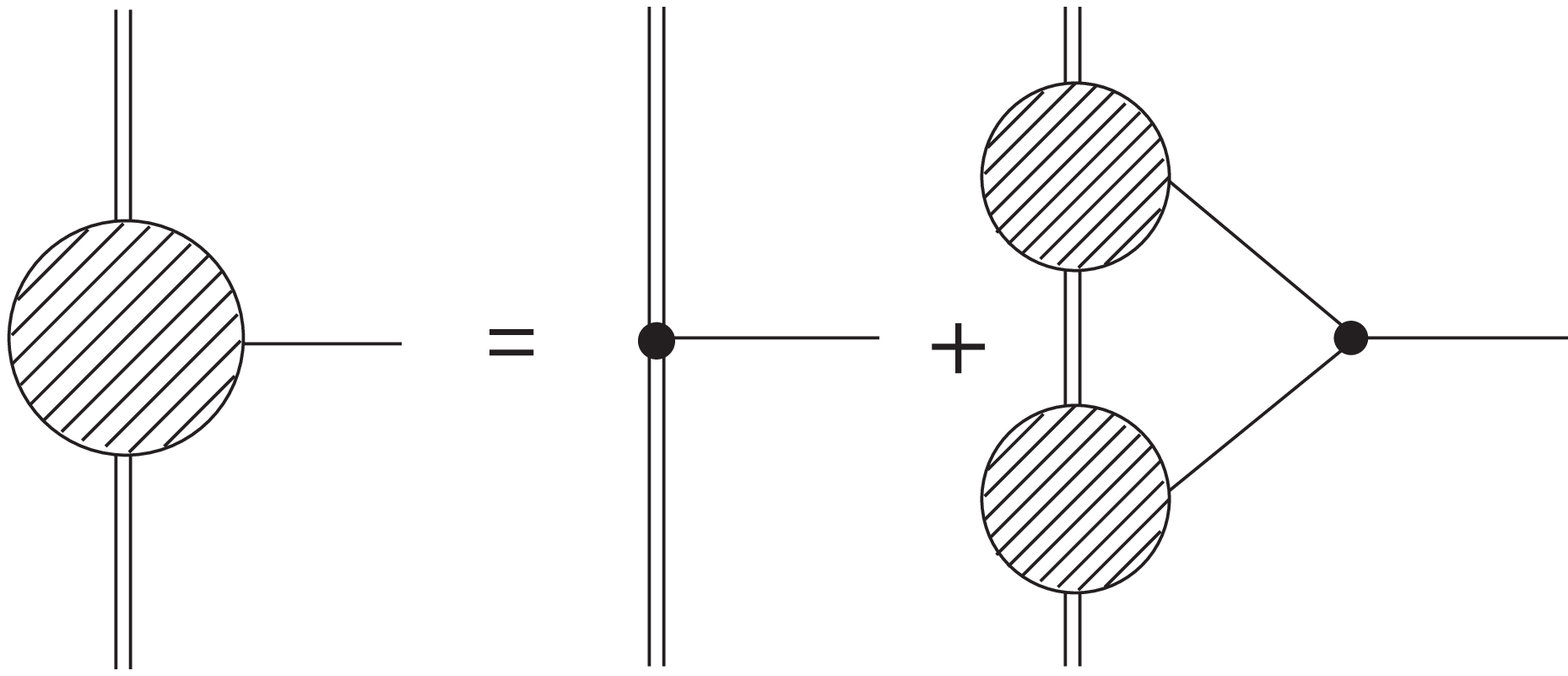}
\caption[]{A schematic diagrammatic representation of the
Schwinger-Dyson recursive integral equation satisfied by the dressed
couplings. More details are given in the body of the paper.}
\label{recursion-schem-fig}
\end{figure}

In addition, the dressing procedure is shown to economize the
calculation in a more significant way as follows. We find that the
dressed couplings satisfy a certain recursive integral equation
schematically described in figure \ref{recursion-schem-fig}. A
perturbative expansion of the solution to this equation equals an
infinite sum of diagrams, and in this sense it encodes many
diagrams. This is nothing but the classical version of the
Schwinger-Dyson equations which were first written in the context
of quantum electro-dynamics \cite{SD}. Yet the current classical
version is of a higher practical value compared to its quantum
counterpart since it does not involve the infinitely many dressed
bulk vertices.

Some discussions of classical versions of the Schwinger-Dyson equations appeared already, yet they all appear to consider a significantly different context. \cite{Blagoev:2001ze} studied
unequal time correlation functions of a non-equilibrium classical field theory, while
\cite{Duetsch:2002yp} aims at giving a construction of the local algebras of observables in
quantum gauge theories.

The paper is divided into two parts. In section
\ref{dress-CLEFT-section} we describe and discuss the dressed
perturbation theory for a general classical (effective) field
theory, while in section \ref{dressing-PN-section} we apply it to
the Post-Newtonian theory. We start in section
\ref{dress-CLEFT-section} by considering a simple scalar classical
field theory coupled to point-like particles as the context for
introducing the required dressing concepts. In subsection
\ref{factorizable-subsection} we characterize the fully
factorizable diagrams before turning in subsection
\ref{dressed-def-subsection} to our main definition, that of the
dressed quantities. In subsection \ref{equivalence-subsection} we
define the dressed perturbation theory and assert that it is
equivalent to the bare theory.

In subsection \ref{integral-subsection} we explain the recursive
integral equation \`{a} la Schwinger-Dyson and comment on its
relation with related concepts. We close the general theory
section in subsection \ref{skeleton-topology-subsection} with a
classification of irreducible skeletons at low loop numbers -
diagrams which are neither factorizable nor do they include
dressed sub-diagrams.

In section \ref{dressing-PN-section} we apply these concepts of
dressing theory to better organize and economize the two-body
Post-Newtonian (PN) effective action. In subsection
\ref{PN-Feynman-subsection} we set up the problem by displaying
the Post-Newtonian effective action and the associated Feynman
rules. In subsection \ref{PNdressed-charges-subsection} we
explicitly compute the three charges in PN. The charge which
couples to the gravitational field is the energy-momentum tensor,
and since the gravitational field naturally decomposes in the PN
limit into three NRG-fields the source decomposes too into three
corresponding parts: the energy density, the momentum density and
the stress. These charges are computed both in $k$-space and in
position space up to the order required to reproduce the 2PN
results, and the stress is determined to one additional order.

In \ref{PNskeletons-subsection} we analyze qualitatively the
diagrams relevant to the computation of the two-body effective
action. Starting with 0PN and making our way to order 2PN we
explain how some of the diagrams can be expressed through dressed
charges, and which diagrams can be generated by the integral
equation. Up to order 2PN only a single (non-tree) diagram is
found to be irreducible. We confirmed that the known 2PN effective
action is reproduced after incorporating our expressions for the
dressed charges from the previous subsection. While thus far
almost all the computations are related to the known
\cite{BlanchetRev} 2PN effective action which was already
reproduced within the EFT approach with NRG fields in
\cite{GilmoreRoss}, here we list all the skeletons required for
the computation of 3PN, thereby indicating the road-map for
organizing the 3PN computation.  In subsection
\ref{computing-beyond-subsection} we convincingly demonstrate the
utility of our method by computing certain novel 3PN and 4PN
diagrams.\footnote{The computed first Post-Minkowskian
approximation \cite{PM-LSB} includes information about a certain
class of diagrams to arbitrarily high PN orders, which are
different however from the ones we compute here.} Finally, in
appendix \ref{master-int} we collect some useful integrals.

\section{Dressed perturbation theory in CLEFT}
\label{dress-CLEFT-section}

In this section we define a dressed perturbation theory in the general CLEFT (Classical Effective Field Theory)  context, namely
for any classical non-linear field theory coupled to point-like sources.

In order to illustrate the main ideas in a simple setting we consider a scalar field
model. The generalization to an arbitrary CLEFT is straightforward and will be spelled out at the end of subsection \ref{dressed-def-subsection} on p. \pageref{gen-def}.

Consider the following bulk action \be
 S_{\rm bulk}[\phi] = \int d^4x\, \[ -\half (\vec{\nabla} \phi)^2+ \frac{1}{2 c^2} \dot{\phi}^2 - \frac{\alpha}{6}\, \phi^3 \] \label{bulk-action}
 \ee
for a scalar field $\phi$ with propagation speed $c$
and coefficient of cubic interaction $\alpha$,
 where $\phi$ couples to any point particle of mass $m$ and charge $q$ through
\be
 S_p = -(m-q) \int d\tau- q \int e^{\phi(x(\tau))}\, d\tau \label{particle-action} ~, \ee
 $d\tau$ is the proper time element and the particles will be assumed non-relativistic $d\vec{x}/d\tau \ll
 c$ (or even static in part of the discussion).
 \emph{The total action} for a many body system is \be
 S = S_{\rm bulk} + \sum_a S_{p,a} ~, \label{full-scalar-action} \ee
 where $S_{p,a}$ is the world-line action (\ref{particle-action}) of the $a$'th particle characterized by $m_a,\, q_a$.

Let us briefly discuss the considerations for choosing this form of the action. The retardation term proportional to $1/c^2$ is considered as a small perturbation in the non-relativistic limit and represents a general small perturbation of the quadratic term.  The term proportional to $\alpha$ was chosen to represent any non-linear interaction.
The interaction term $- q \int \exp\(\phi(x(\tau)) \)$ includes the charge coupling $- q \int \phi(x(\tau))$ together with some
representative non-linear terms (this exponential form of the interaction appears in Post-Newtonian theory, for example).

The bulk theory (\ref{bulk-action}) has a vacuum at $\phi=0$ and we consider the
perturbation theory around it. As usual, we note that while this
vacuum is unstable, it could be stabilized by adding a mass term
to $S_{\rm bulk}$, and it could even be made a global minimum
through the addition of a quartic term in the potential.

In this paper we study the \emph{two-body problem} rather than the
seemingly more general \emph{many body problem} since it is the
simplest and currently the most interesting case. The
generalization to the many body problem seems straight forward.

\begin{figure}[t!]
\centering \noindent
\includegraphics[width=10cm]{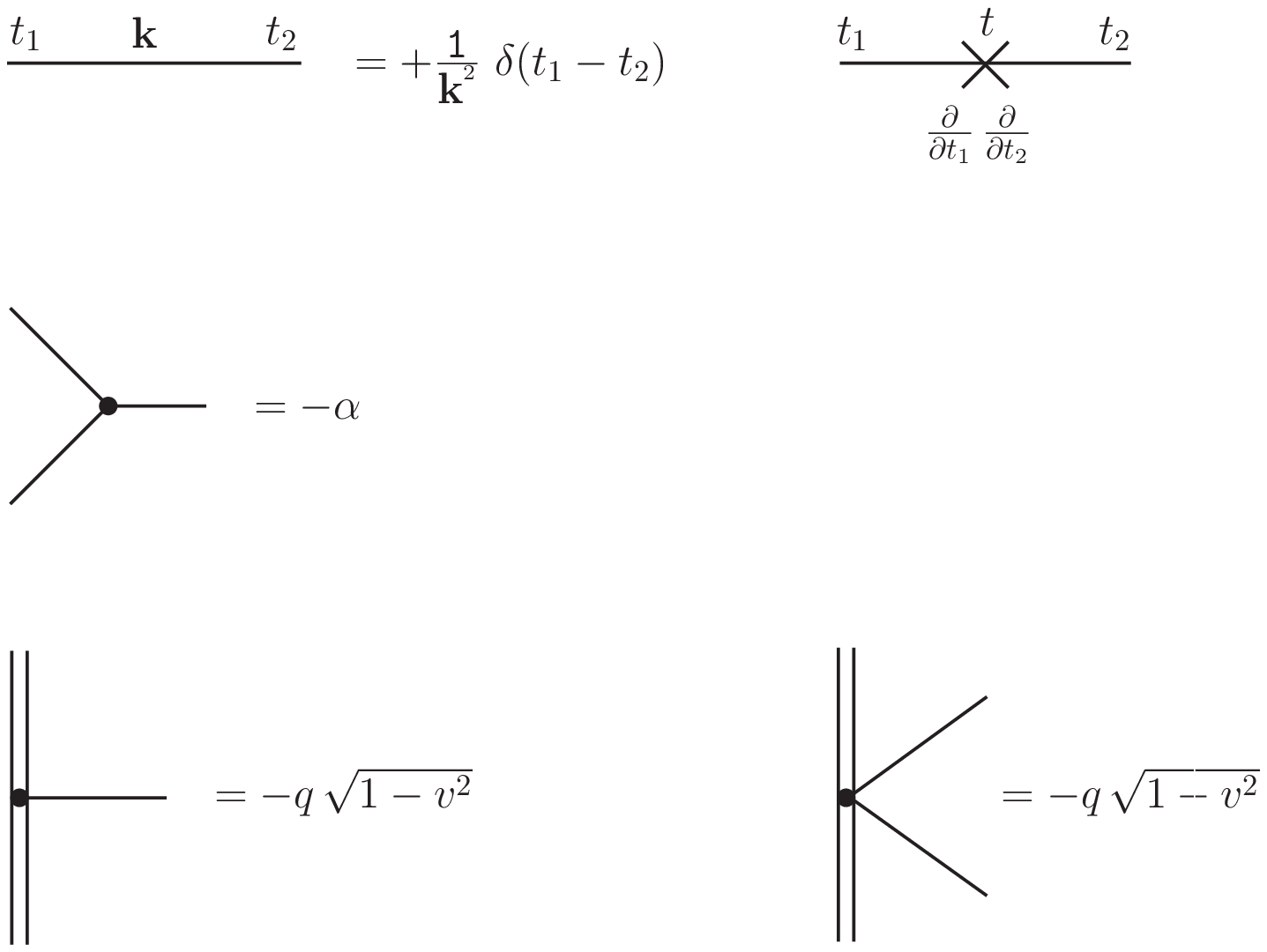}
\caption[]{The Feynman rules involving $\phi$ for the static
scalar field theory whose action is given by
(\ref{full-scalar-action}). The rules describe: the propagator and
the quadratic perturbation vertex -- retardation (top), the cubic
bulk vertex (middle), and the world-line vertices (bottom) where
the ellipsis stand for additional non-linear world-line vertices.
Hereafter $k$ denotes a spatial wave-number.}
\label{scalar-Feynman-rules-fig}
\end{figure}

The \emph{Feynman rules} involving $\phi$ are shown in figure
\ref{scalar-Feynman-rules-fig}. They are standard except
for the CLEFT conventions which makes them real -- vertices are read from the
action and propagators are given schematically by $-1/S_2$ where
$S_2$ is the quadratic part of the action. See \cite{CLEFT-caged}
for the original definition and full detail.

The two body effective action is defined by \be
 S_{eff}[x_1,x_2] := S[x_1,x_2,\phi(x_1,x_2)] \ee
where the RHS represents the action (\ref{full-scalar-action})
evaluated on the solution $\phi=\phi(x_1,x_2)$ given the particle
trajectories $x_1(\tau_1),\, x_2(\tau_2)$. The effective action is
known to be equal to the sum of all connected Feynman diagrams
made out of $\phi$ propagators with arbitrary bulk vertices and
world-line vertices but without loops of propagating fields (such
classically forbidden loops are allowed quantum mechanically).

\subsection{Factorizable diagrams}
\label{factorizable-subsection}

There are several possible paths to classical dressed perturbation theory. We shall build it from
first principles and later discuss its relations with both the quantum version and the standard classical theory.

The main idea is to economize the perturbation theory by identifying certain recurring sub-diagrams. Before we proceed to the more essential, dressed sub-diagrams we discuss a stronger and simpler form of reduction, namely factorizable diagrams.

\begin{figure}[t!]
\centering \noindent
\includegraphics[width=4cm]{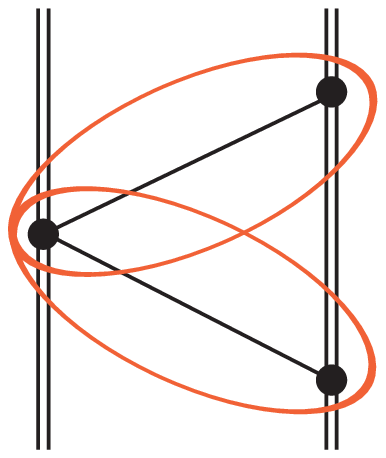}
\caption[]{An example of a factorizable diagram (the simplest).
The two factors are circled (in red) and they intersect in a NL-WL
vertex. A diagram of this type appears at 1PN -- see figure
\ref{1PN-fig}.} \label{factorizable-fig}
\end{figure}

A Feynman diagram is called \emph{factorizable} whenever the
expression which it represents factorizes (into a product of
factors), each one corresponding to a sub-diagram. An example is
shown in figure \ref{factorizable-fig}. In CLEFT we have the
interesting property that \emph{a diagram of the 2-body effective
action is factorizable if and only if it contains a Non-linear
world-line (NL-WL) vertex}, where a NL-WL vertex is a vertex with
more than a single bulk field.

Indeed, if a diagram contains a NL-WL vertex then since quantum
loops are not allowed, each leg of the vertex generates a separate
sub-diagram. Conversely, a connected\footnote{All the diagrams of
the 2-body action are connected by definition.} factorizable
diagram necessarily factorizes at a vertex. A bulk vertex would
not serve since wave-number conservation couples between all its
legs.\footnote{In CLEFT we cannot have a factorizable diagram such
as the ``figure 8'' diagram in $\phi^4$ theory, since we cannot
have $k$ conservation for any strict subset of propagators leaving
the bulk vertex, as each propagator connects to a distinct
world-line vertex (or vertices) where $k$ is arbitrary.}

\subsection{Dressed charge and propagator}
\label{dressed-def-subsection}

Following our discussion of factorization we proceed to consider only diagrams without any NL-WL vertices.
Any non-factorizable diagram contains various sub-diagrams. We wish to further economize the perturbation theory by identifying in a unique and natural way a class of sub-diagrams which repeatedly show up
at high orders of the two body effective action. We shall call them ``dressed sub-diagrams''.
We start with constructive definitions and an explanation of the name's origin, to be followed by a more abstract characterization through an equivalence of perturbation theories which serves to explain the rational behind the definitions.

\begin{figure}[t!]
\centering \noindent
\includegraphics[width=15cm]{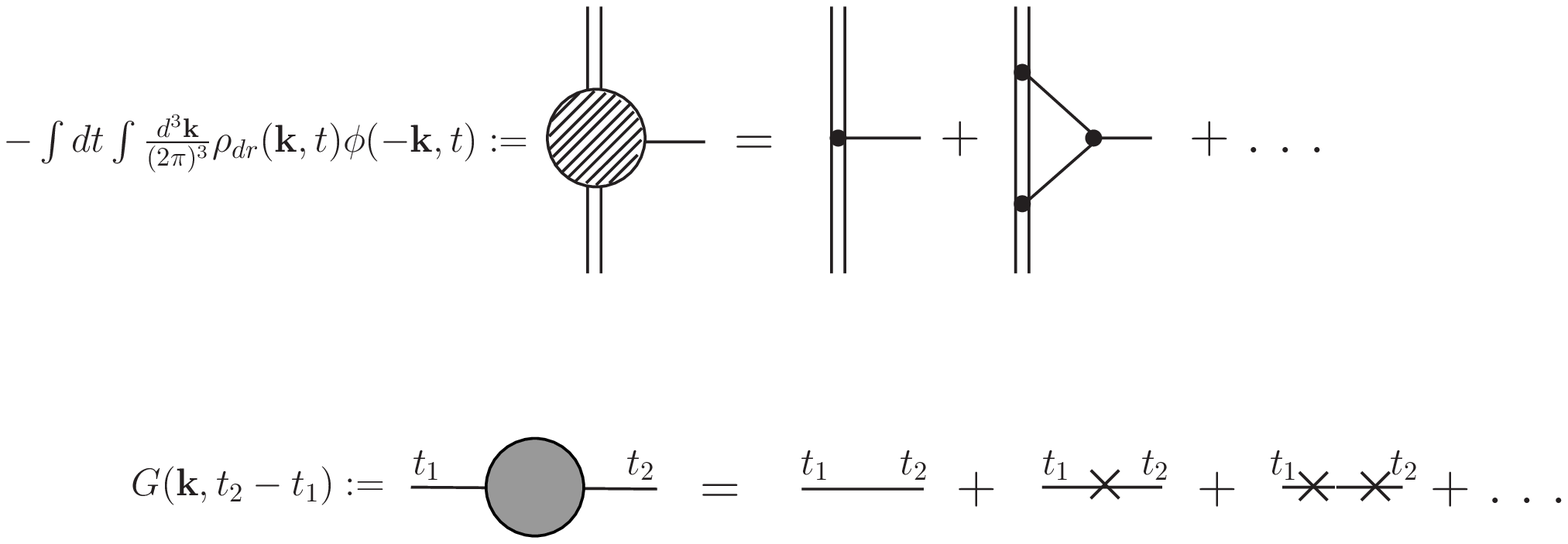}
\caption[]{The diagrammatic definition of the dressed charge
distribution (top) as the one point function for $\phi$ in the
presence of a single source and the dressed propagator (bottom) as
the full two point function for $\phi$ in vacuum. Retardation
vertices are forbidden on the external leg of the dressed charge.}
\label{dressed-subdiag-fig}
\end{figure}

\presub {\bf Definitions} \bi
 \item The \emph{dressed charge of the particle}, $\rho_{dr}(k,t)$
is defined diagrammatically through figure
\ref{dressed-subdiag-fig} (top). The dressed charge is an infinite
sum of sub-diagrams, where each summand will be called a
\emph{dressed charge sub-diagram}.

\item The \emph{dressed propagator}, $G_{dr}(k,\Delta t)$
 is defined diagrammatically through figure \ref{dressed-subdiag-fig} (bottom).
 Each summand in the definition will be called a \emph{dressed propagator sub-diagram}.

\ei

\begin{figure}[t!]
\centering \noindent

\includegraphics[width=4cm]{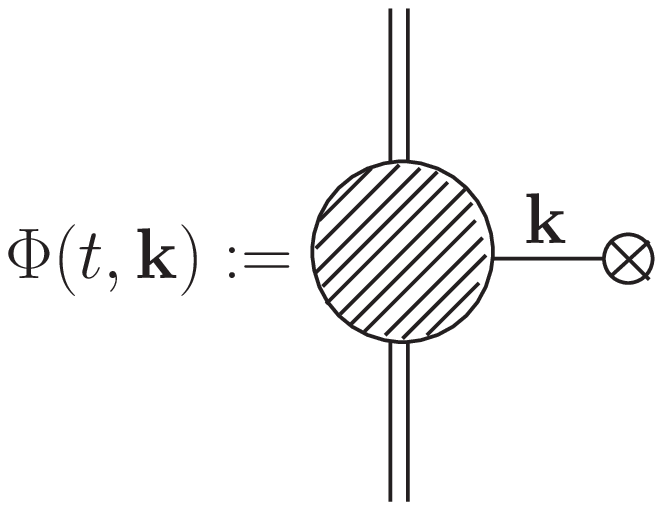}
\caption[]{The diagrammatic representation for the value of the field in the presence of the point-particle source.
The only difference with the definition of the dressed charge in figure \ref{dressed-subdiag-fig} (top) is an added propagator on the external leg.}
\label{phi-one-point-fig}
\end{figure}

Let us inspect these definitions.
 In equations $\rho_{dr}(r)$ is defined through \be
 \rho_{dr}(r,t) := \triangle \Phi(r,t) = \int d\tau\, q\, \delta^{(4)}\(x-x(\tau)\)  + \half \alpha\, \Phi^2 +\del_t^2 \Phi \label{Gnqdr} \ee
where the second equality is a differential equation which
together with retarded boundary conditions defines $\Phi(r,t)$,
the full solution for the field $\phi$ in presence of the given
source world-line. This equation is equivalent to the diagrammatic
definition since $\Phi$ is given diagrammatically by figure
\ref{phi-one-point-fig},
 and hence in
$k$-space $\Phi(k,t)=-\rho_{dr}(k,t)/k^2$ from which (\ref{Gnqdr})
follows in position space.

The dressed charge describes the apparent particle charge
(distribution) at long distances. It arises from the non-linear
interactions of the scalar field which ``dress'' the point charge,
and is useful for studying the dynamics of a system composed of
several such particles. The term which we use, ``dressed'' charge
(and propagator) is standard terminology in quantum field theory
(see for example \cite{AlkoferSmekal}) as well as in classical
field theory. For completeness we would like to mention some
related terms.  The first is renormalization, where like here one
replaces a bare quantity by a scale-dependent renormalized
quantity which is defined through the divergences of the same
sub-diagrams.\footnote{The idea of renormalization in the PN
context appeared already in the literature. \cite{DimReg}
associated renormalization with the regularization of certain
divergences which appear in the particle's effective action at
order 3PN. \cite{GoldbergerRothstein1} studied the same
divergences from the EFT approach and in particular they computed
the renormalization of the energy-momentum tensor of a static
particle up to 2 loops. Later \cite{CannellaSturani} also studied
the renormalization of the energy-momentum tensor.}
 Another related term is resummation where one discusses partial sums of
diagrams.

The equation form of the dressed
propagator is \bea
 G_{dr} &=& \frac{1}{\Box}=\frac{1}{k^2+\del_t^2/c^2}=\frac{1}{k^2}\, \frac{1}{1+\del_t^2/(c^2 k^2)} = \non
        &=& \frac{1}{k^2}\, \( 1 - \frac{1}{c^2\, k^2} \del_t^2 + \frac{1}{c^4\, k^4} \del_t^4 +\cdots \) \label{Gren} \eea
where the last equality is a representation of the series on the
left of figure \ref{dressed-subdiag-fig} (bottom), while the
previous expressions can be considered to be a closed form
summation of that series.

Physically, in this field theory and also in PN the dressed propagator is nothing but
 the fully relativistic propagator. However, this need not be the case in general.

\presub {\bf Example}. To illustrate these ideas let us discuss an
example. A detailed application to the Post-Newtonian theory will
be given in section \ref{dressing-PN-section}.

\begin{figure}[t!]
\centering \noindent
\includegraphics[width=9cm]{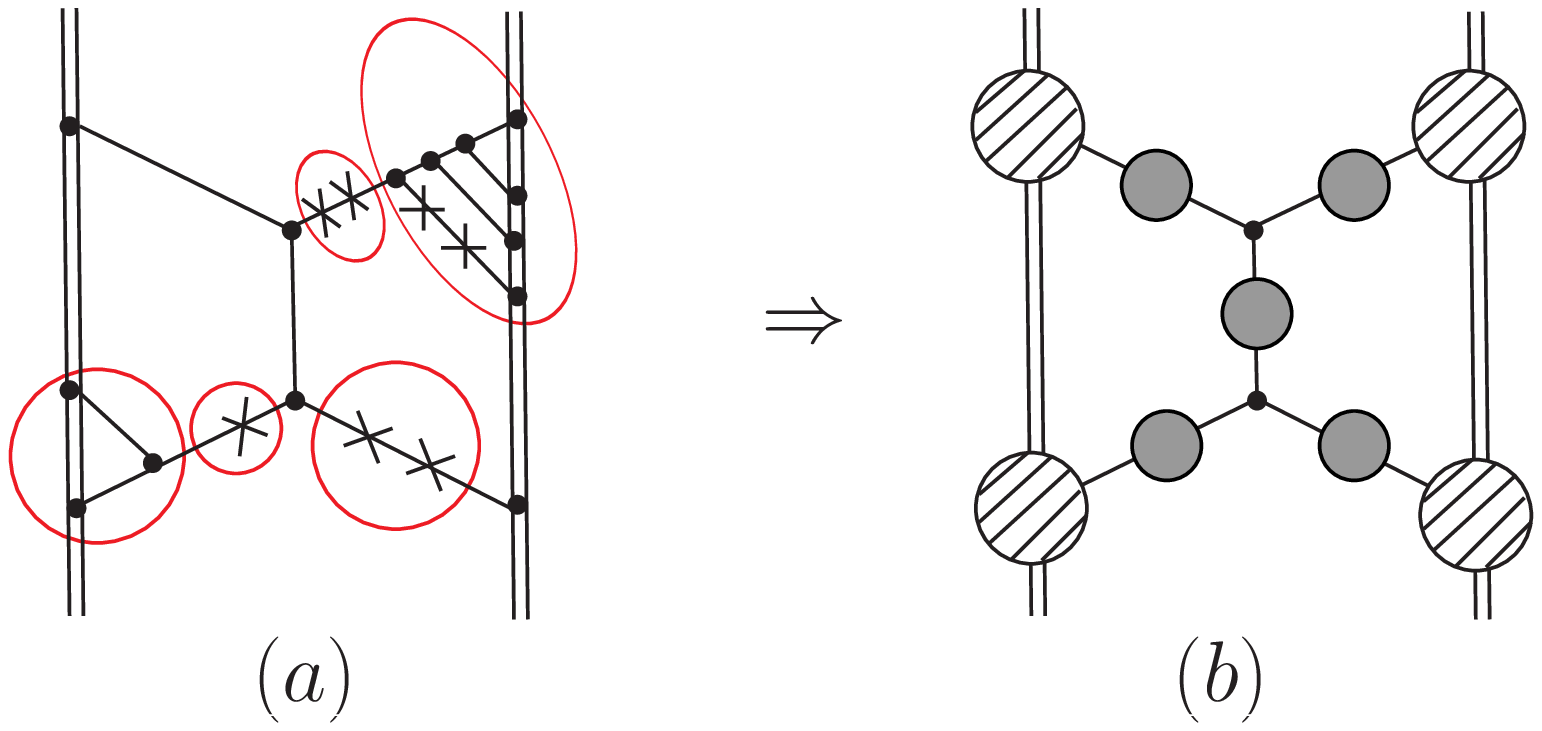}
\caption[]{An example for the correspondence between diagrams of
the bare theory, such as (a) and the corresponding dressed
diagram (b). Note a standard subtlety: retardation insertions are allowed inside a dressed charge vertex,
but not on its external leg. See text for further discussion.} \label{example-fig}
\end{figure}

Consider the 6-loop Feynman diagram in figure \ref{example-fig}(a)
which contributes in the bare theory to the 2-body effective
action. It has five dressed sub-diagrams (circled) -- two dressed
charges and three dressed propagators. Note that all the
world-line vertices of each dressed charge belong to {\it one and
the same} point particle. Note also that retardation vertices are
allowed inside the dressed charge but not on its external leg.
Upon replacing the dressed sub-diagrams by dressed vertices and
propagators we obtain the diagram's skeleton\footnote{This term
will be further discussed and defined in the next subsection.} in
(b). The resulting diagram is only two-loop. The other loops were
absorbed by the dressed charges. Note that the 3-loop dressed
charge sub-diagram includes in it other dressed sub-diagrams, but
these are not maximal.

\presub {\bf Generalization}. So far we worked in a simple setting
of a cubic scalar field model. Here we shall indicate how the
definitions given above for the dressed sub-diagrams generalize to
a general field theory. \label{gen-def}

In the scalar field theory we had a single dressed vertex and
a single dressed propagator. In a general Classical field
theory interacting with point-like particles we have a
dressed charge for each field. For the dressed charge to be non-trivial the bulk theory must
be interacting.

The dressed propagator in a general field theory is labeled by any
two fields that can ``mix''. In case the fields do not mix then we
have a single dressed propagator for each field. For the dressed
propagator to be non-trivial we need the quadratic Lagrangian to
decompose into a leading part and a perturbation, so that the
leading part will determine the bare propagator, while the small
part will determine the two-vertex. The dressed propagator will
then be proportional to the inverse of the full quadratic
Lagrangian. We note that one may choose to diagonalize the dressed
propagator and accordingly redefine the fields such that there
will be no mixing through dressed propagators.

\subsection{Equivalence of perturbation theories}
\label{equivalence-subsection}

We proceed to define a dressed perturbation theory, which is
equivalent to the original one, but somewhat more economic.

\presub {\bf Definition}. A Feynman diagram which includes a
non-trivial dressed sub-diagram of the form shown in figure
\ref{dressed-subdiag-fig} (\emph{dressing sub-diagram}) will be
called \emph{dressing-reducible}. Otherwise it will be called
\emph{dressing-irreducible}.

\begin{figure}[t!]
\centering \noindent
\includegraphics[width=6cm]{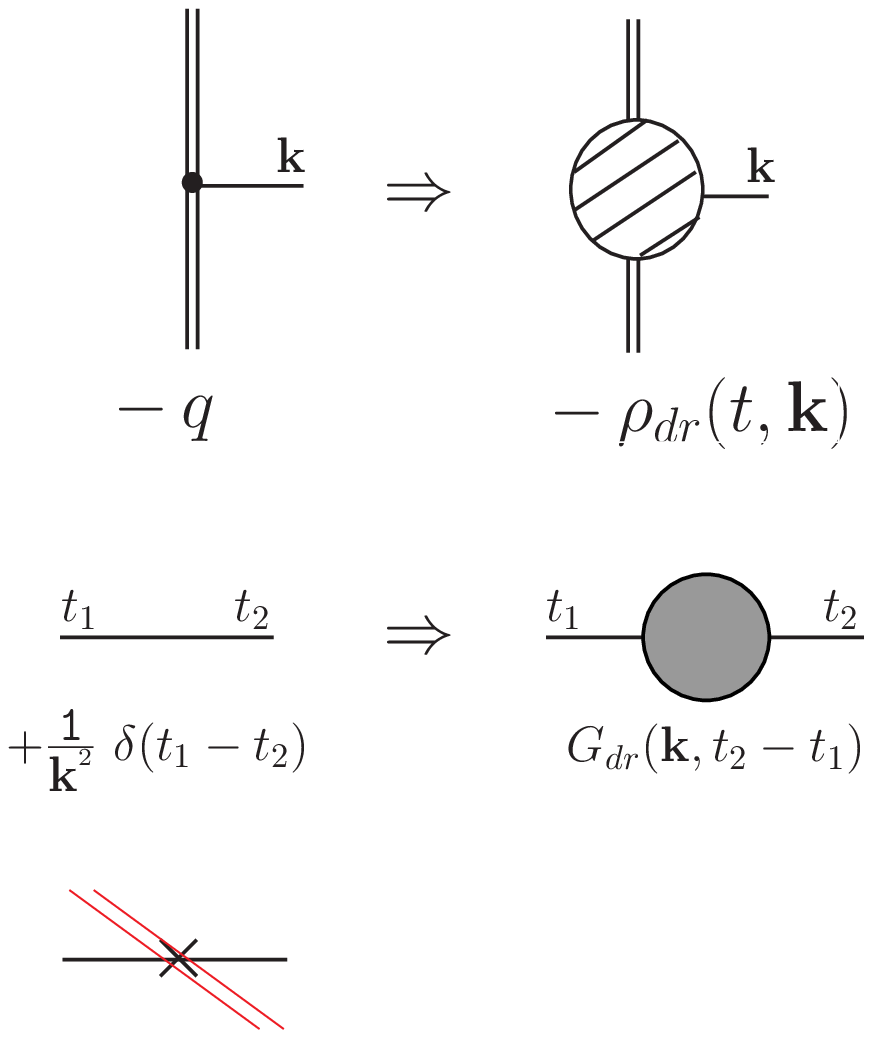}
\caption[]{The dressed Feynman rules contain changes relative to
the bare Feynman rules in figure \ref{scalar-Feynman-rules-fig}.
The bare world-line vertex (charge) Feynman rule changes into the
dressed one, the bare propagator changes into the dressed one, and
the retardation rule is omitted. The rest of the rules remain
unchanged.} \label{dressed-Feynman-rules-fig}
\end{figure}

\presub {\bf Definition}. The \emph{dressed perturbation
theory} is defined as follows \bi
 \item Figure \ref{dressed-Feynman-rules-fig} shows the changes in Feynman rules
  relative to the original theory (figure \ref{scalar-Feynman-rules-fig}).
 \item Only dressing-irreducible diagrams are allowed.
 \ei

The original perturbation theory will be distinguished from the
dressed one by referring to it as \emph{bare}.

\presub {\bf Property} -- \emph{The bare and dressed
perturbation theories are equivalent}: each diagram of the bare theory is included exactly once in a dressed diagram.
 \vspace{.5cm}

This property is analogous to a standard one holding for dressed
actions in QFT, and we shall outline a proof. Given a bare diagram
we claim that its maximal dressed sub-diagrams are unique. We find
the uniqueness property to be quite apparent when one thinks about
it, but we shall not attempt to provide here a proof, as it seems
tedious. Given the decomposition we replace the dressed
sub-diagrams by propagators and vertices of the dressed theory
according to figure \ref{dressed-Feynman-rules-fig}. The resulting
reduced diagram is called \emph{the skeleton} of the original
diagram. The uniqueness of decomposition implies now that each
bare diagram is included once and only once in the dressed
perturbation theory, and hence the two are equivalent.

\presub {\bf Discussion}. Before proceeding to
describe another property of the dressed theory, namely,
the integral equation, let us pause to discuss some aspects of the
definitions and property above.

\emph{Rational behind definitions}. The decomposition into a skeleton with blobs
which represent dressed sub-diagrams is natural in a practical, computational
sense. When computing a diagram, the dressed sub-diagrams
are almost inevitably evaluated on the way, by their nature. Hence it is computationally
natural to prepare a list of dressed sub-diagrams and
their value, in order to avoid their repeated evaluation, which is
exactly what the dressed theory does.

Yet, the computational argument alone does not fix our definitions
of the dressed sub-diagram. \emph{The characterizing property is precisely
the equivalence property above}, namely that the dressed perturbation theory is equivalent to the bare one,
or in other words, \emph{the dressed theory is self-sufficient or autonomous}. We claim (again without proof) that this property can be used to \emph{derive} our definitions.

 \emph{Analogy with effective action in Quantum Field
Theory (QFT)}. The ideas above are analogous to those of the
effective action in standard QFT. There one collects all the
1-particle-irreducible (1PI) diagrams into the effective action
and then allows in the perturbative expansion only tree diagrams
of the effective action. In both cases it is important that the
decomposition is unique -- any diagram can be uniquely divided
into 1PI sub-diagrams which allow a reduction to a tree skeleton.

\emph{Relation of the dressed sub-diagrams with the ``dressed
source'' in standard QFT}. The two notions are essentially the
same. Note however that unlike some cases, for us it is important
that the dressed charge sub-diagrams are only those where all
world-line vertices belong to one and the same point particle.

Our notion of the dressed propagator is also essentially the same as
the dressed propagator and the associated field strength renormalization in standard QFT.
The 2-vertex in CLEFT is the ``self energy'' (the 1PI two point
function), and (\ref{Gren}) is essentially the standard QFT
relation between the self energy and the dressed propagator.
The difference is that in QFT many diagrams can contribute to the
self energy (actually normally they are infinitely many
corresponding to an arbitrary number of possible loops) while in
CLEFT the 2-point vertex is read directly from the Lagrangian.

\begin{figure}[t!]
\centering \noindent
\includegraphics[width=8cm]{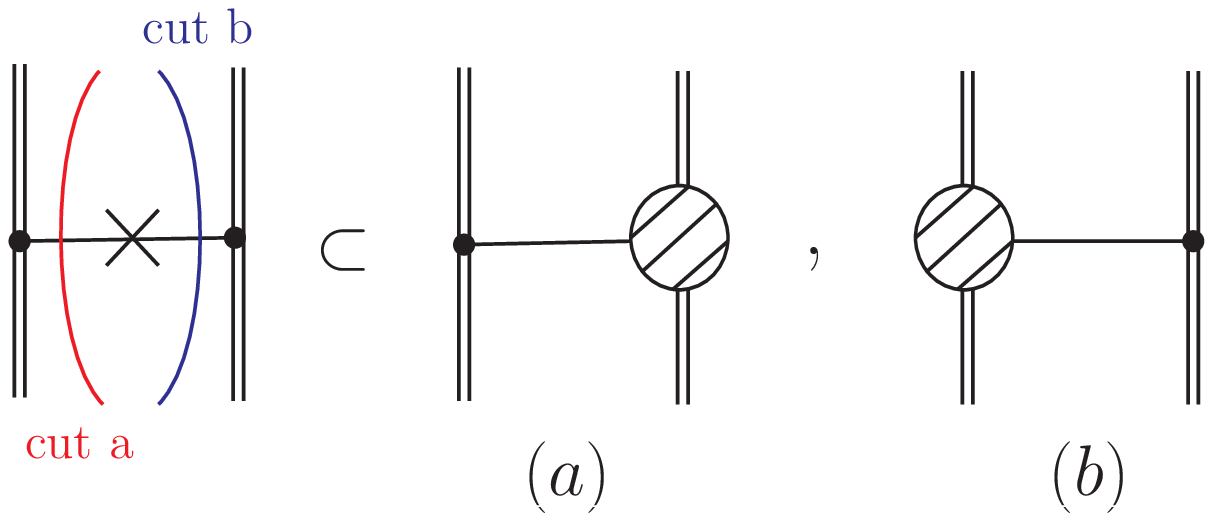}
\caption[]{An over-counting problem with an alternative definition
of the (non-static) dressed perturbation theory, as discussed in
the text.} \label{counter-expl-fig}
\end{figure}

\emph{Why is the dressed propagator necessary?} Consider
doing away with the definition of the dressed propagator by
allowing the dressed vertex sub-diagrams to include
retardation vertices on the external leg. In the classical theory this would actually work
in all but one important class of diagrams, that of the
dressed Newtonian interaction -- figure \ref{0PN-renorm-fig}, where
the correspondence between the bare and dressed diagrams
would break. For example, consider the diagram in figure
\ref{counter-expl-fig}.
There are two distinct ways to
cut the diagram into sub-diagrams, the cuts being denoted by a,b.
Accordingly this diagram is doubly counted in the putative
dressed theory, thereby disqualifying it. This is all the better
since the dressed propagator is an appealing physical concept
which we would not want to lose anyway.

\subsection{The Schwinger-Dyson recursive integral equation in CLEFT}
\label{integral-subsection}

In this section we shall describe a second property of the dressed perturbation theory: certain
recursive relations which generally take the form of integral equations. We start by considering the
static limit of the scalar theory (\ref{full-scalar-action}) where the idea is simpler to illustrate and later we refine it to include the general non-static case.

Recall the definition of $\rho_{dr}(r)$ in figure
\ref{dressed-subdiag-fig}.

\begin{figure}[t!]
\centering \noindent
\includegraphics[width=9cm]{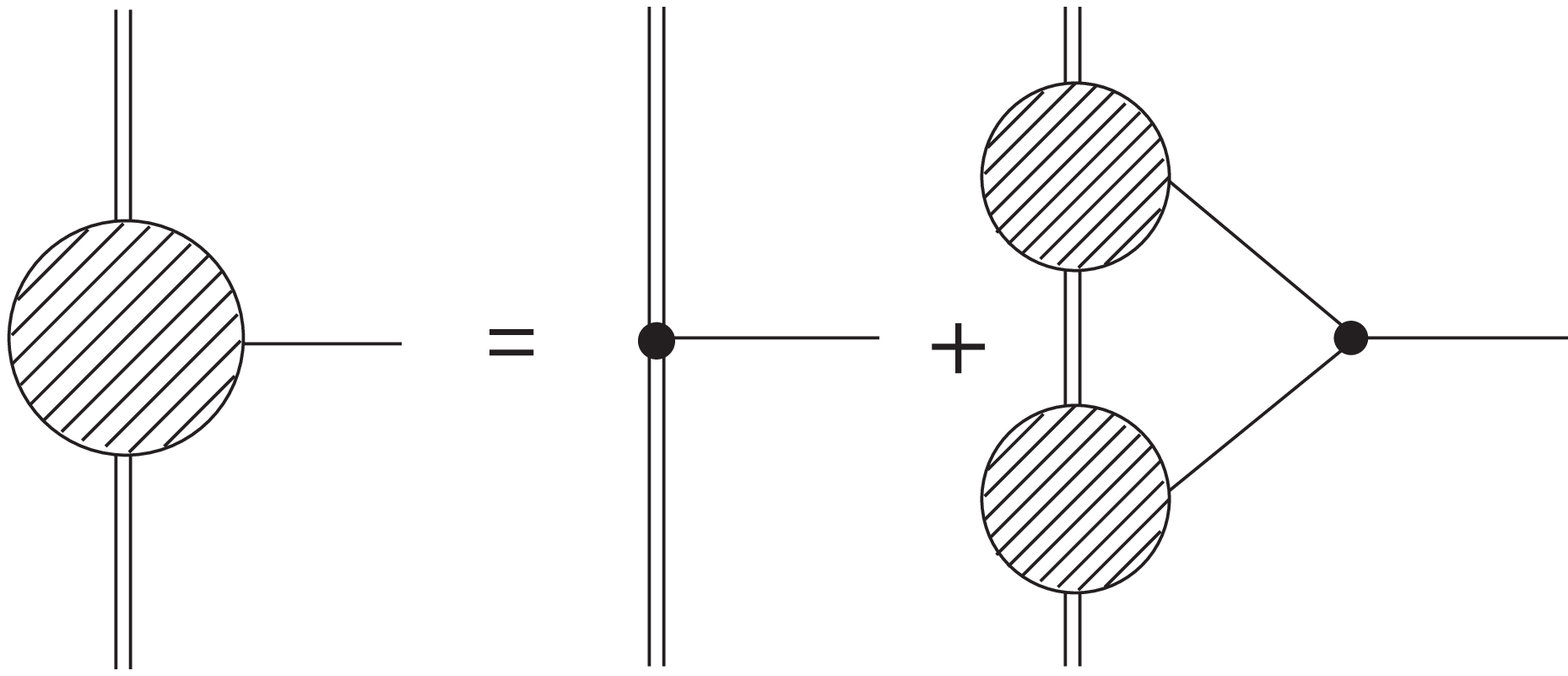}
\caption[]{The diagrammatic representation of the recursive
integral equation satisfied by the dressed charge in the static
limit.} \label{vertex-recursion2-fig}
\end{figure}

\presub {\bf Property}. \emph{The dressed quantities satisfy recursive relations.}
 In the static limit $\rho_{dr}(r)$ satisfies the recursive relation which is
shown diagrammatically in figure \ref{vertex-recursion2-fig}. Its
equation form reads \be
 -\int \frac{d^3k}{(2 \pi)^3} \rho_{dr}(k) \phi(-k) =
 -\left. q\, \phi\right|_{\vec{r}=0} - \frac{\alpha}{2} \int \frac{d^3k}{(2 \pi)^3} \int \frac{d^3k_1}{(2 \pi)^3}
 \frac{\rho_{dr}(k_1)}{k_1^2} \frac{\rho_{dr}(k-k_1)}{(k-k_1)^2}  \phi(-k) ~.\ee
 After factoring out $\int d^3k\, \phi(-k)/(2 \pi)^3$
 we are left with the following integral equation for $\rho_{dr}(k)$ \be
   \rho_{dr}(k) = q +  \frac{\alpha}{2} \int \frac{d^3k_1}{(2 \pi)^3}\,  \frac{\rho_{dr}(k_1)}{k_1^2}\, \frac{\rho_{dr}(k-k_1)}{(k-k_1)^2}   ~. \label{integ-eqn-charge} \ee
 \vspace{0.5cm}

Given a small $\al$ the integral equation can be solved
perturbatively in $\al$ by expanding $\rho_{dr}(k)= \sum \al^n
\rho^{(n)}(k)$. The zeroth order is given by $\rho^{(0)}(k)=q$,
the first order is given by $\rho^{(1)}(k)=\half  \int d^3k_1/(2
\pi)^3\, \alpha\, \frac{q}{k_1^2}\, \frac{q}{(k-k_1)^2}$ and so
on. This iterative solution of the integral equation is precisely
equivalent to the diagrammatic expansion in figure
\ref{dressed-subdiag-fig} (because they both compute the same
quantities, namely $\rho^{(n)}(k) )$.

The advantage of the integral equation over the diagrammatic
expansion is that it is shorter and economizes the computation by
avoiding the need to identify all the necessary diagrams and
compute them. Some readers may benefit from the following analogy:
the relation between the recursive relation and the full
diagrammatic expansion is analogous to the relation between
defining a function through a differential equation and defining
it through the corresponding power series which solves the
differential equation.

\begin{figure}[t!]
\centering \noindent
\includegraphics[width=10cm]{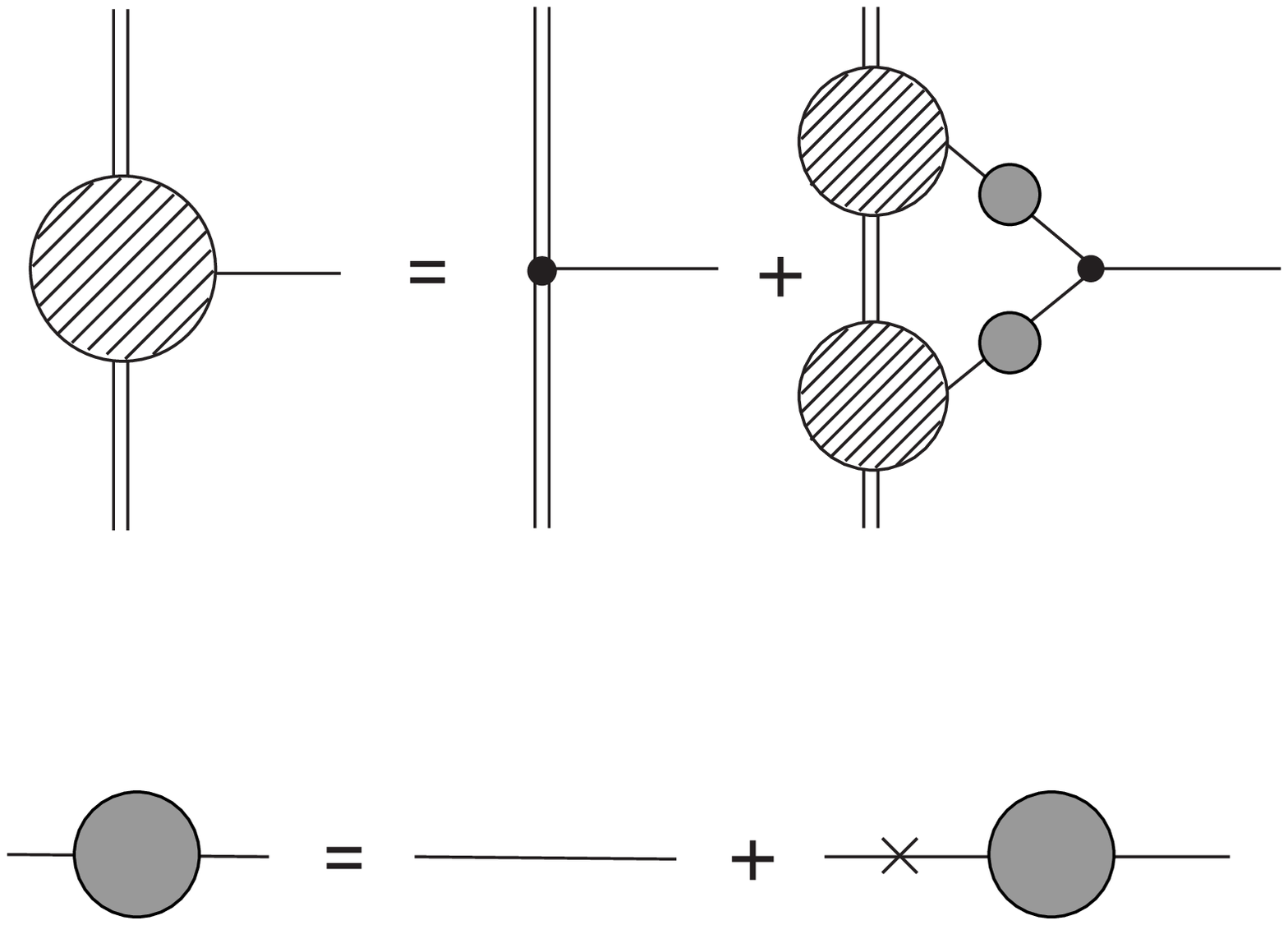}
\caption[]{A diagrammatic representation of the full set of (the two) recursive equations satisfied by the dressed charge (top) and propagator (bottom) in the general, non-static case. Note a change in the recursive relation for the dressed charge relative to the static equation in figure \ref{vertex-recursion2-fig}.}
\label{recursion-fig}
\end{figure}

The non-static case. Once we restore time dependence into the
action (\ref{full-scalar-action}) we expect the
dressed propagator to have a role in the recursive relations
as well. First, it has a recursive
 relation of its own given diagrammatically
by figure \ref{recursion-fig} (bottom). In this case the recursive
relation is actually algebraic rather than integral, and hence the
closed form solution (\ref{Gren}) exists. Secondly, the recursive
relation for the dressed vertex (figure
\ref{vertex-recursion2-fig}) is refined to include the dressed
propagator and is now given by figure \ref{recursion-fig} (top).

We now proceed to make several comments.

\bi

\item The recursive relation (figures
\ref{vertex-recursion2-fig},\ref{recursion-fig} and equation
\ref{integ-eqn-charge}) can be considered as inherited from its
quantum version (see for example \cite{BjorkenDrell}), but there
are interesting differences. As it is, figure
\ref{vertex-recursion2-fig} does not hold in \emph{quantum} field
theory \footnote{except for the tree diagram limit of course.}
because there can be additional $\phi$ propagators connecting the
two blobs on the RHS. Since such additional propagators would
create a closed loop of propagating fields it is forbidden in
CLEFT. In order for the recursive integral relations to apply in
the quantum case one must generalize them to include the dressed
bulk vertices yielding the celebrated Schwinger-Dyson equations
\cite{SD}. However, as stated in several textbooks the practical
usefulness of the quantum version is rather limited since more and
more Green's functions with more external legs participate in the
equations as the order is increased
 \footnote{``From the point of view of making a practical calculation we have accomplished little; the unknown quantities ... have been expressed
 in terms of yet another unknown...'' \cite{BjorkenDrell},
  ``... the system involves an infinite hierarchy of equations ... their usefulness is limited'' \cite{ItzyksonZuber}.}
  while in CLEFT the equations are more practical exactly because there are only a finite number of dressed quantities.

\item The dressed mass $\rho_{dr}(r)$ is closely related to the
field profile $\Phi(r)$ generated by the point-like source.
Indeed, each one of them is sufficient to determine the other
through (\ref{Gnqdr}). While $\rho_{dr}(r)$ satisfies an integral
equation, $\Phi(r)$ satisfies a ``mirror'' differential equation,
namely the equation of motion given by the second equality in
(\ref{Gnqdr}). The perturbative expansion of the integral equation
is dual in turn to the perturbative expansion of the differential
equation into some sort of a power series (which will generally
include $\log$ factors as well).

\item Relation with the beta function. The recursive integral
equation (\ref{integ-eqn-charge}, and figures
\ref{vertex-recursion2-fig},\ref{recursion-fig}) determines
$\rho_{dr}(k)$ and so does the beta function. Yet, the two
equations are different as the beta function is a first order
differential equation for $d\rho_{dr}/d\log(k)$. Therefore it must
be that the beta function equation is a special or limiting case
 (whose precise definition will not be pursued here)
 when the leading behavior of $\rho_{dr}$ is logarithmic in
$k$.
\ei

\subsection{Irreducible 2-body skeletons}
\label{skeleton-topology-subsection}

\begin{figure}[t!]
\centering \noindent
\includegraphics[width=9cm]{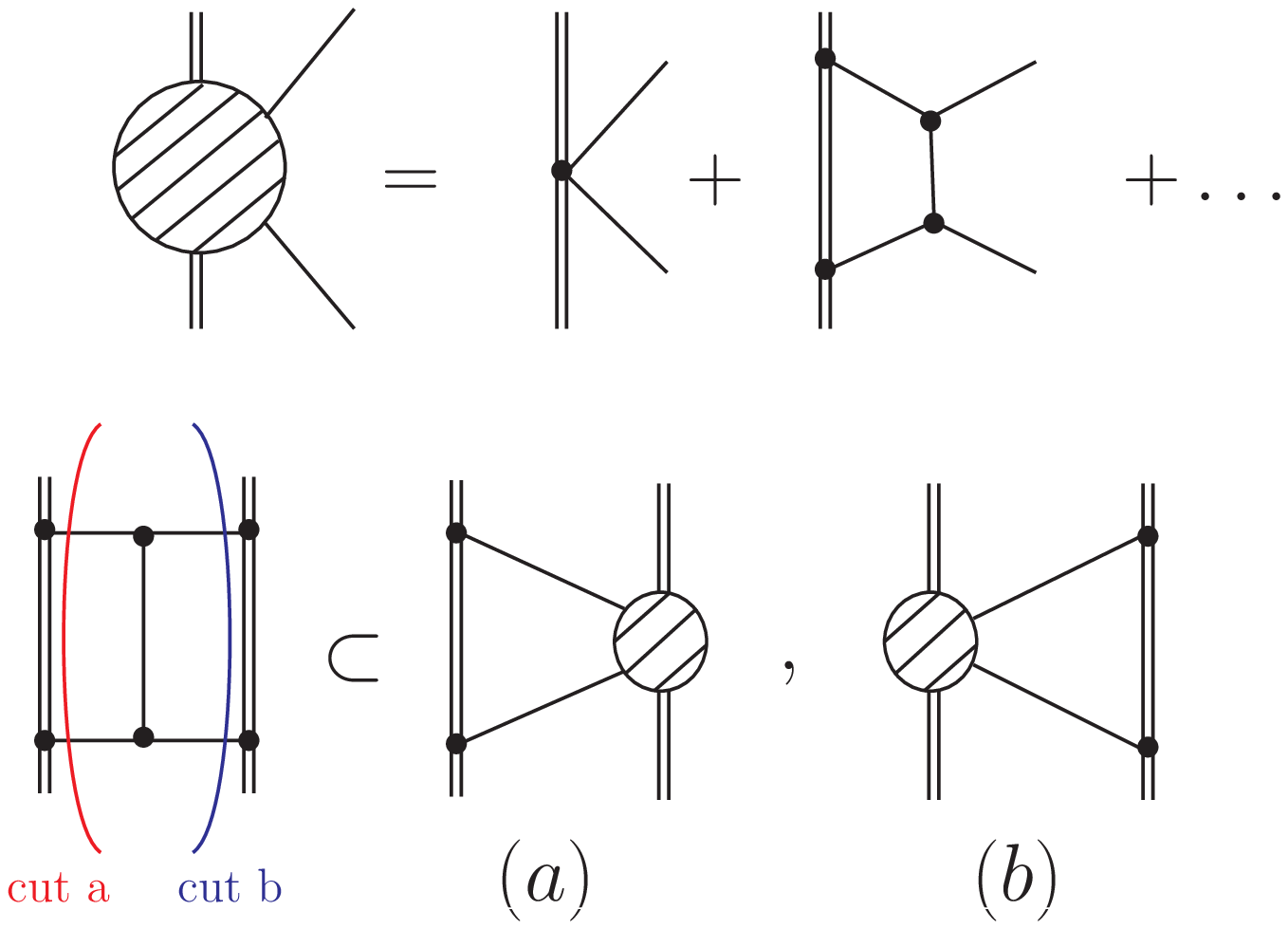}
\caption[]{The top line shows a possible definition for a 2-field
dressed world-line vertex. The second line demonstrates the
over-counting problem which occurs. The bare diagram on the left
can be cut into sub-diagrams in two different ways denoted (a) and
(b).} \label{multiple-field-renorm-fig}
\end{figure}

Consider the non-linear world-line vertices (NL-WL vertices) in
the scalar action (\ref{full-scalar-action}), namely vertices with more than a single bulk
field.
Our definition of the dressed charge concerns a world-line
vertex with a single bulk field. It is possible to generalize the
definition of the dressed vertex to any number of bulk
fields, such as the 2-field vertex shown on the top line of figure
\ref{multiple-field-renorm-fig}. In this way one may define a
\emph{fully dressed one body effective action}.

However, for the purpose of computing the 2-body effective action
\emph{we use only the dressed charges (and propagators) and for
the NL-WL vertices we use the bare vertices rather than the the
dressed ones} because they would have created a problem: the
decomposition of diagrams would not have remained unique (just as
in our discussion around figure \ref{counter-expl-fig}). For
example, the diagram on the second row of figure
\ref{multiple-field-renorm-fig} can be decomposed in two {\it
different} ways by the two shown cuts, and that corresponds to
doubly counting this diagram in the putative fully dressed
perturbation theory.

Recall (subsection \ref{factorizable-subsection}) that non-linear world-line vertices are interesting for another,
complementary property:
\emph{a diagram of the 2-body effective action is factorizable if and only if it contains a Non-linear world-line vertex}.
From this perspective the above-mentioned issue with using dressed WL-NL vertices
is all the better since there is no need for dressing --
all diagrams with NL-WL vertices are factorizable and hence reduce
to computations of lower order and lower loop number.

\begin{figure}[t!]
\centering \noindent
\includegraphics[width=9cm]{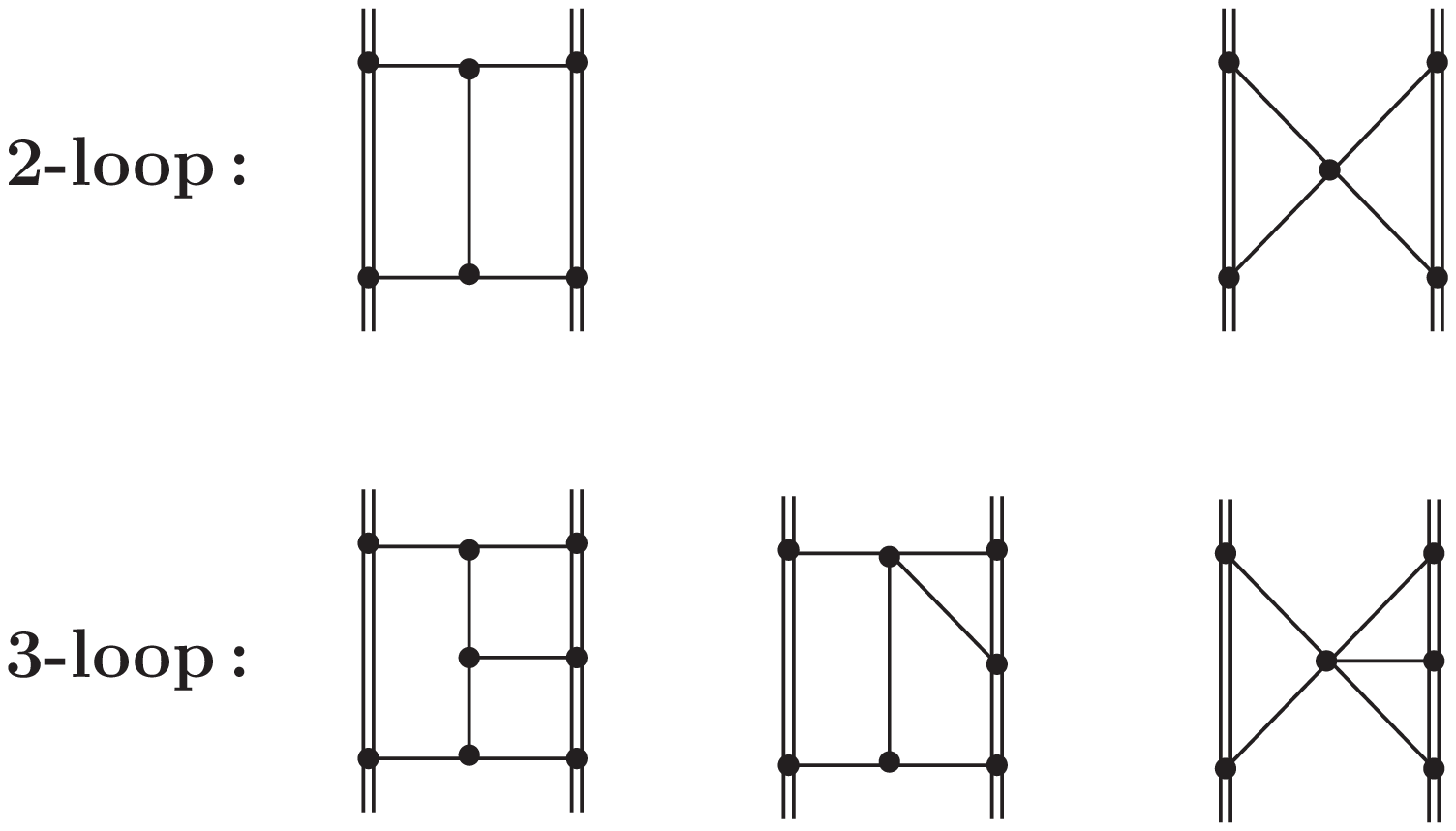}
\caption[]{Classification of topologies for irreducible diagrams
of the 2-body effective action, both non-factorizable and
dressing-irreducible. The top line shows the possibilities at
2-loop, and the bottom line shows the possibilities at 3-loop.}
\label{irred-top-fig}
\end{figure}

It is interesting to \emph{classify the possible irreducible
diagrams in the 2-body effective action} -- those which are both
non-factorizable and dressing-irreducible. The possible topologies
are independent of the details of the specific field theory. At
1-loop there are no irreducible diagram topologies. At 2-loop (top
line of figure \ref{irred-top-fig}) there is the ``H'' diagram on
the left and its degeneration - the ``X'' diagram where the two
cubic bulk vertices degenerate into a quartic vertex. At 3-loop
(bottom line of figure \ref{irred-top-fig}) there is the topology
shown on the left, together with its two possible degenerations.
We did not list here the trivial 0-loop (tree-level) diagram shown
in figure \ref{0PN-fig}.

\section{Dressing the 2-body Post-Newtonian problem}
\label{dressing-PN-section}

In this section we apply the dressed perturbation theory to the
Post-Newtonian (PN) expansion, thereby demonstrating its utility.
We start in subsection \ref{PN-Feynman-subsection} by setting up
the problem, establishing the conventions, and displaying the
Post-Newtonian effective action and the associated Feynman rules.
In subsection \ref{PNdressed-charges-subsection} we explicitly
compute the three PN charges both in $k$-space and in position
space up to the order required to reproduce the 2PN results, apart
from the stress that is determined to one additional order. In
\ref{PNskeletons-subsection} we analyze qualitatively the diagrams
relevant to the computation of the two-body effective action
including a list of all the skeletons required for the computation
of 3PN. In subsection \ref{computing-beyond-subsection} we compute
certain novel 3PN and 4PN diagrams.

\subsection{Effective action and Feynman rules}
\label{PN-Feynman-subsection}

Consider a binary system composed of slowly moving massive objects.
Replacing the vicinity of each of these objects by the relativistic
point particle coupled to gravity and neglecting the effects of the
higher-dimensional tidal terms, leads to the following effective
action
 \be
 S=S_{EH}+S_{pp} \, .
 \ee

In PN the gravitational metric field is
naturally divided into three fields of
Non-Relativistic Gravity (NRG-fields): $\phi$ the Newtonian
potential, $A_i$ the gravito-magnetic vector potential and
$\gamma_{ij}$ a symmetric tensor (with spatial indices)
\cite{NRG}. The field redefinition $g_{\mu\nu} \to (\phi,A_i,\gamma_{ij})$
is defined by \cite{CLEFT-caged,NRG,GilmoreRoss}
 \be
 ds^2 = g_{\mu\nu}\, dx^\mu\, dx^\nu
      = e^{2 \phi}\, \(dt-A_i\, dx^i\)^2 - e^{-2\phi/(d-3)}\, \gamma_{ij}\, dx^i dx^j ~,
 \label{KKansatz}
 \ee
 where we take the space-time dimension $d$ to be arbitrary and only at the end we shall specialize to 4d.
The bulk action is the Einstein-Hilbert action expressed in terms of NRG-fields  \be
 S_{EH}[g] = \frac{1}{16\pi G}\int R[g] \rightarrow S_{EH}[\gamma,A,\phi]
 ~.
 \label{eqn:EH_KKdecomp}
 \ee

The point particle trajectory is denoted by $\vec{x}=\vec{x}(t)$ and we shall denote by
$\vec{v}=\dot{\vec{x}}(t)$ and $\vec{a}=\ddot{\vec{x}}(t)$ its 3-velocity and 3-acceleration, respectively. The point particle action is given by
 \bea
 S_{pp} &=& -\sum_a m_a \int d\tau = -\sum_a m_a \int dt\, e^\phi\,
 \sqrt{(1-\vec{A} \cdot \vec{v}_a)^2-e^{-2(d-2)\phi/(d-3)}\, \gamma_{ij}\,
 v_a^i\, v_a^j } = \non
 &=& -\sum_a m_a \int dt\, \sqrt{1-v_a^2}
 \non
 &-& \sum_a m_a \int dt\, \(\frac{(d-3) + v_a^2}{ (d-3)\sqrt{1-v_a^2} } \, \phi
 - {\vec{A} \cdot \vec{v}_a\over \sqrt{1-v_a^2}} - {\sigma_{ij}\, v_a^i\, v_a^j\over 2 \sqrt{1-v_a^2}}
 + \dots \) \, ,
 \label{pp-in-D}
 \eea
where the dummy index $a$ runs over all the masses involved in the
binary evolution, in the second equality equation (\ref{KKansatz})
was applied, the ellipsis denote terms non-linear in the bulk
fields $(\phi,A,\gamma)$ and we define
 \be
 \sigma_{ij} := \gamma_{ij} -\delta_{ij} ~.
 \ee

The Einstein-Hilbert action is invariant under reparametrizations
and should be gauge fixed. Leaving aside the question
which gauge would be optimal for this problem,  we choose the
harmonic gauge in order to facilitate comparison with the
literature. Accordingly, we add the following gauge fixing term to
the Einstein-Hilbert action
 \be
 S_{GF}={1 \over 32 \pi G} \int d^{\,d}x \sqrt{g} ~
 \Gamma^{\mu}\Gamma^{\nu} g_{\mu\nu} ~,
 \label{gauge-fixing} \ee
where $\Gamma^{\mu}=\Gamma^{\mu}_{\alpha\beta} g^{\alpha\beta}$.

\begin{figure}[t!]
 \centering \noindent
 \includegraphics[width=7cm]{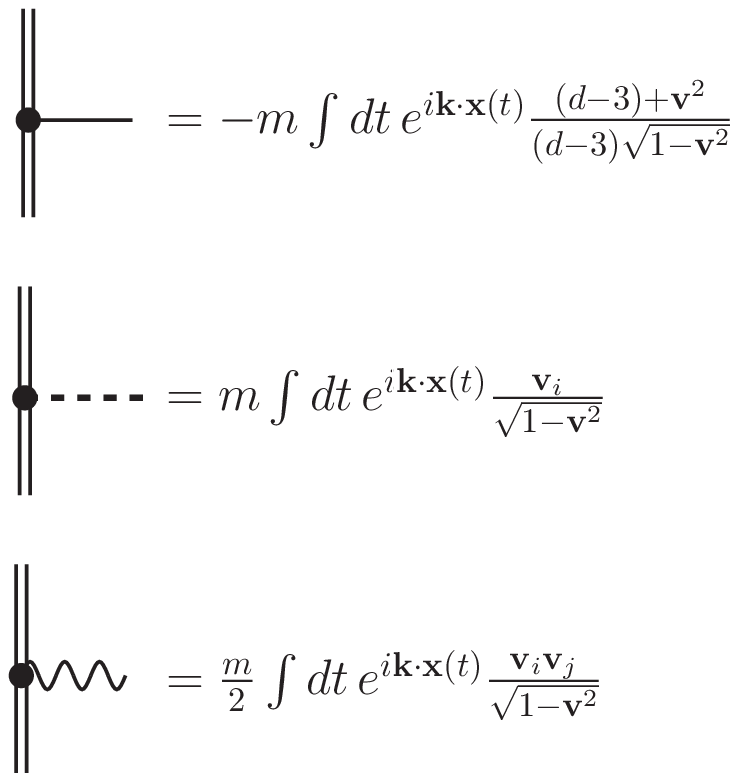}
 \caption[]{Feynman rules obtained from the expansion of (\ref{pp-in-D}) up to linear order in $\phi$, $A_i$ and $\sigma_{ij}$.
 The undetermined wave-numbers flow into the vertex.}
 \label{PN-Feynman-rules-1}
 \end{figure}

The Feynman rules for $\phi$, $A_i$ and $\sigma_{ij}$ coupled
to the world-line can be read from (\ref{pp-in-D}) and we list on
figure \ref{PN-Feynman-rules-1} those couplings which are necessary to
our discussion. Solid, dashed and wavy lines of the figure are
associated with propagators of the instantaneous non-relativistic
modes $\phi$, $A_i$ and $\sigma_{ij}$, respectively. In momentum
space these propagators are given by
\begin{center}
  \begin{picture}(51,77) (50,-109)
    \SetWidth{0.7}
    \SetColor{Black}
    \Line(-37,-40)(3,-40)
    \Text(5,-47.4)[lb]{\Black{$ ~ = \, 8 \pi G  ~ \frac{d-3}{d-2} \, \delta(t-t') ~ \frac{1}{\textbf{k}^2}$}}
    \Line[dash,dashsize=6](-37,-70)(3,-70)
    \Text(-37,-65)[lb]{\Black{$i$}}
    \Text(3,-67)[lb]{\Black{$j$}}
    \Text(5,-77.4)[lb]{\Black{$ ~ = \, - 16 \pi G \, \delta(t-t') \, \delta_{ij} \, \frac{1}{\textbf{k}^2}$}}
    \Photon(-37,-100)(3,-100){4}{4}
    \Text(-37,-95)[lb]{\Black{$i \, j$}}
    \Text(-4,-93)[lb]{\Black{$k \, l$}}
    \Text(5,-107.4)[lb]{\Black{$ ~ = \, 32 \pi G \, \delta(t-t') \, P_{ij,kl} \, {1 \over \textbf{k}^2 }$ }}
  \end{picture}
\end{center}
with $P_{ij,kl}={1 \over
2}\[\delta_{ik}\delta_{jl}+\delta_{il}\delta_{jk}-{2 \over
d-3}\delta_{ij}\delta_{kl}\]$.

\vspace{.2cm}

\begin{figure}[t!]
 \centering \noindent
 \includegraphics[width=14cm]{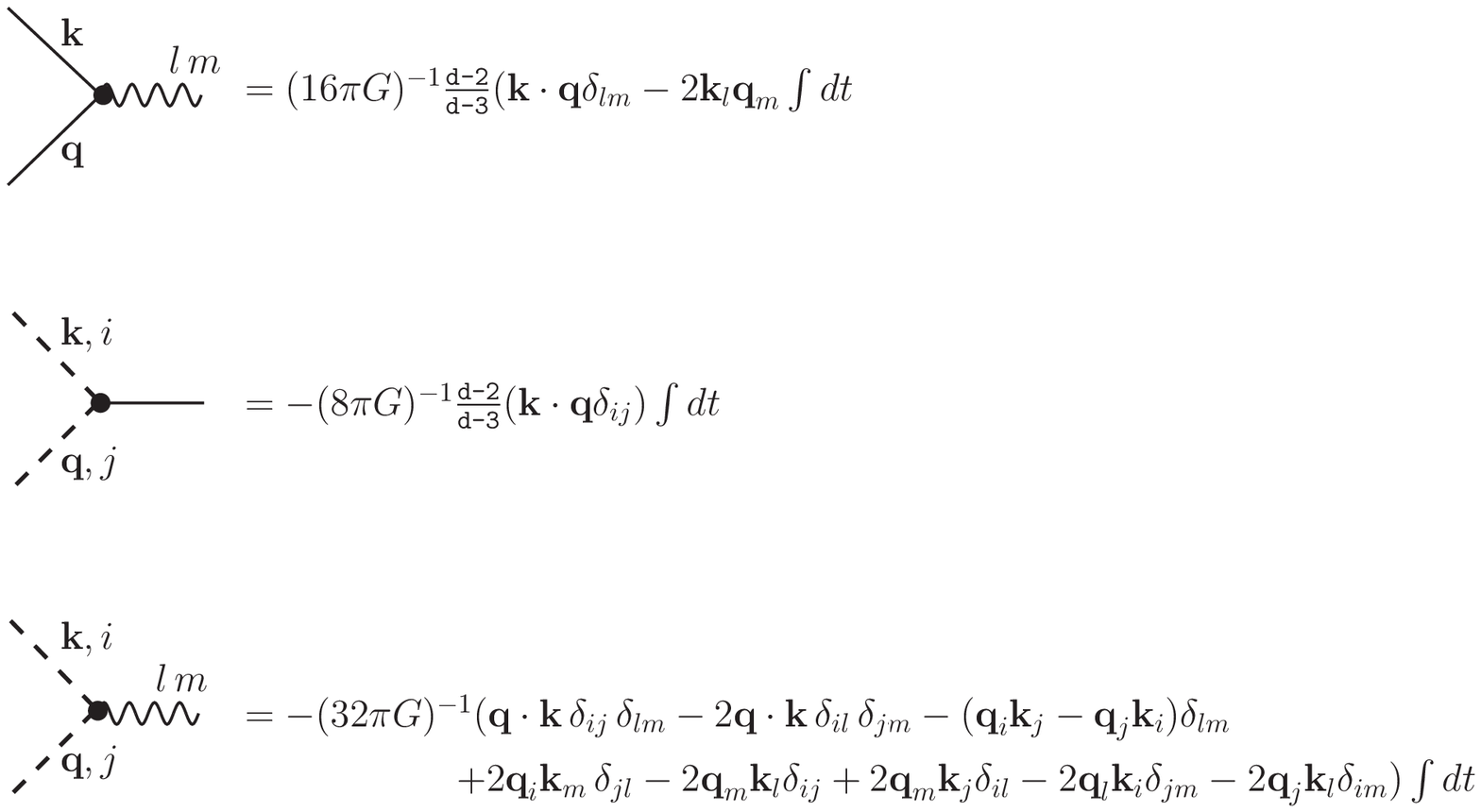}
 \caption[]{Static bulk vertices obtained from the expansion of the
 Hilbert-Einstein action (\ref{eqn:EH_KKdecomp}) and gauge fixing term
 (\ref{gauge-fixing}) but restricting to time-independent terms. The undetermined wave-numbers flow into the vertex.}
 \label{PN-Feynman-rules-2}
 \end{figure}

\begin{figure}[t!]
 \centering \noindent
 \includegraphics[width=14cm]{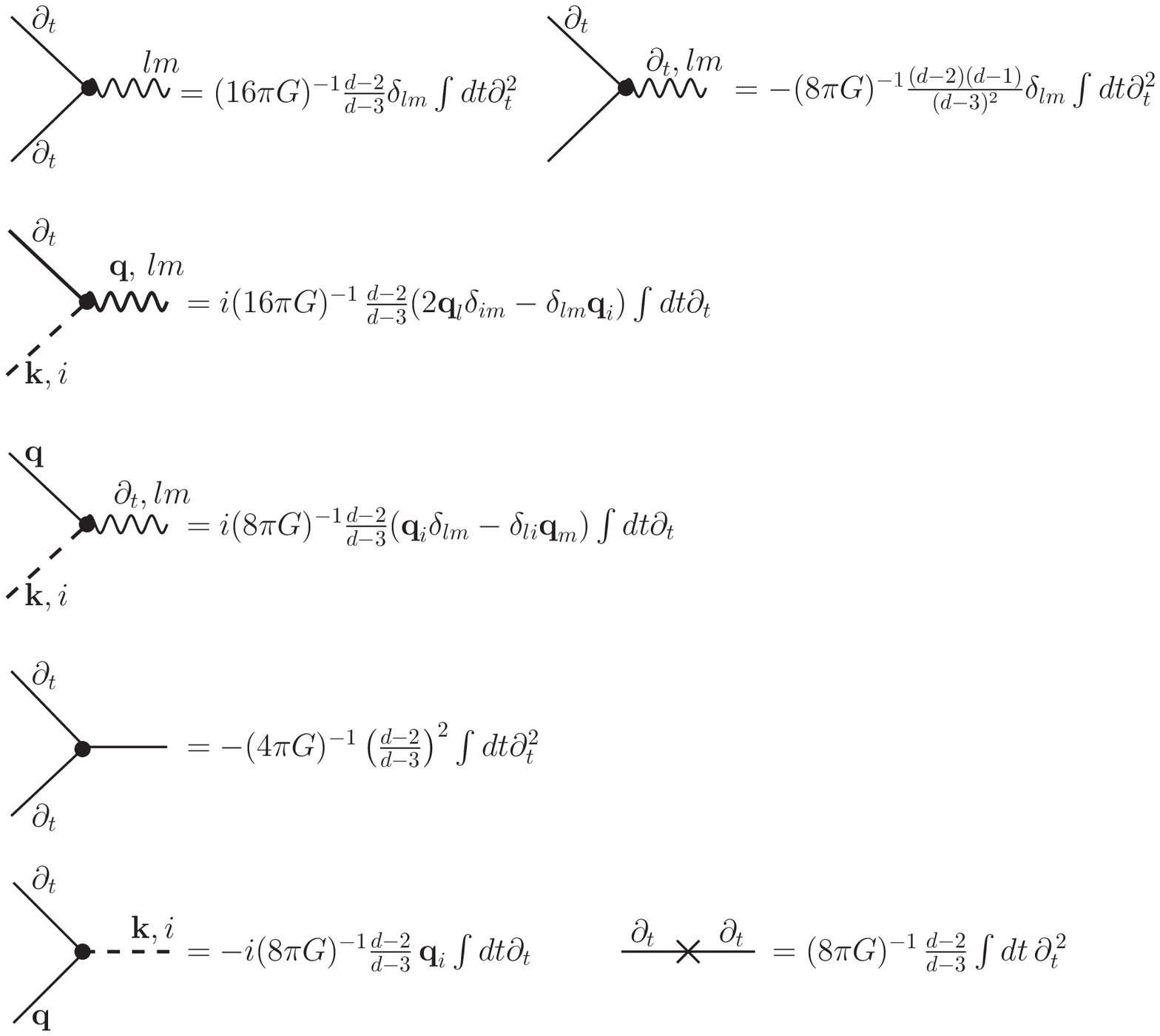}
 \caption[]{Time-dependent bulk vertices obtained from the expansion of the
 Hilbert-Einstein action (\ref{eqn:EH_KKdecomp}) and gauge fixing term
 (\ref{gauge-fixing}). Time derivative above the propagator indicates
 the direction on which it acts. The undetermined wave-numbers flow into the vertex.}
 \label{PN-Feynman-rules-3}
 \end{figure}

The Feynman rules for the bulk vertices are obtained
from expansion of the Einstein-Hilbert action (\ref{eqn:EH_KKdecomp}) and the gauge fixing
term (\ref{gauge-fixing}). The resulting set
of vertices which contribute to the calculations below are presented
in figure \ref{PN-Feynman-rules-2} and figure \ref{PN-Feynman-rules-3},
where we separate vertices which involve time derivatives in figure \ref{PN-Feynman-rules-3}, from the static ones in
figure \ref{PN-Feynman-rules-2}.

In addition, one has to impose  wave-number\footnote{Quantum Mechanically the wave-number $k$ is equivalent to momentum, and this is how $k$ is usually referred to, but in Classical field theory it is not a momentum.} conservation at each bulk vertex by assigning it a delta-function
 factor, $(2\pi)^{d-1}\delta(\sum_i
\textbf{k}_i)$, where $\sum_i
\textbf{k}_i$ is the total wave-number flow into a given vertex, and one must
integrate over each undermined wave-number $\textbf{k}$ of the diagram
 \be
 \int_{\textbf{k}}:=\int {d^{d-1} \textbf{k} \over (2\pi)^{d-1}} ~ .
 \ee
Finally, one has to divide by the symmetry factor of the diagram.

\subsection{Dressed charges}
\label{PNdressed-charges-subsection}

Given the general theory, it is natural to inquire about the form of the dressed quantities within PN.
 Accordingly we would like to compute the dressed charges (the dressed propagators are simple and are also discussed below).

\begin{figure}[t!]
\centering \noindent
\includegraphics[width=5.5cm]{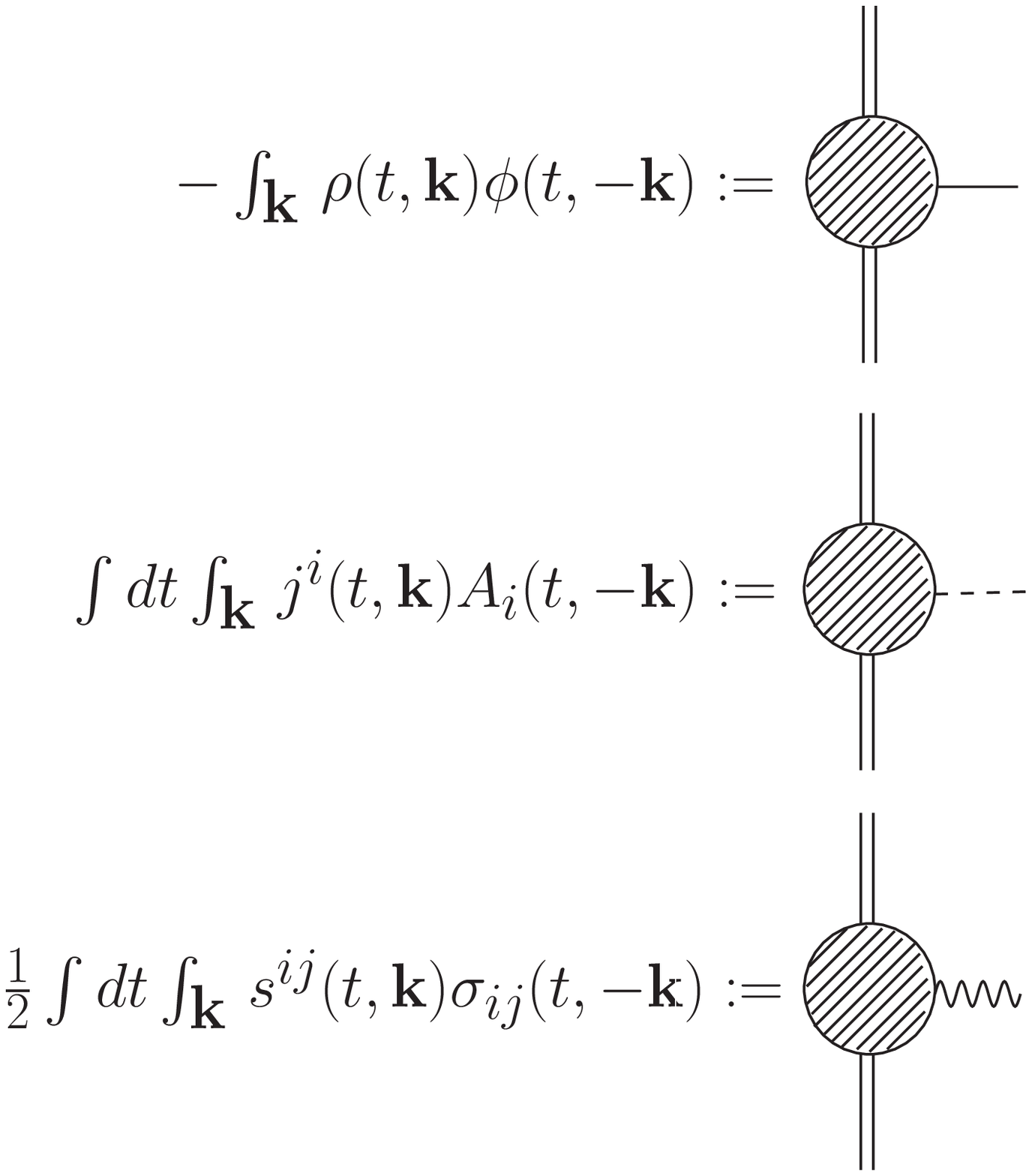}
\caption[]{The diagrammatic definition of the dressed charge
distributions in PN $\rho,j^i$ and $s^{ij}$ as the one point
function for $\phi,\,A_i$ and $\sigma_{ij}$ respectively in the
presence of a single source. Retardation vertices are forbidden
from the external leg.} \label{PN-charges-fig}
\end{figure}

Following our discussion at the end of subsection \ref{gen-def} on
the generalization of the definitions for dressed sub-diagrams to
a multi-field field theory we define three dressed charges for the
interaction of gravity with a compact point-like object. The
$\phi$ charge is usually referred to as energy, the $A$ charge is
the energy current (or alternatively, momentum distribution) and
the $\sigma$ charge is the stress. Together they describe the full
energy-momentum tensor (in space-time). In analogy with figure
\ref{intro-renorm-fig} we define the 3 dressed PN charges in
figure \ref{PN-charges-fig}. All the dressed charges describe the
changing apparent charge due to non-linear bulk interaction at
large, but not infinite, distances.

In PN we have a quadratic retardation vertex for all 3 fields just
like in our scalar field example. Since the 3 fields have
different tensor characters they cannot mix, and we have exactly 3
dressed propagators each one defined as in figure
\ref{dressed-subdiag-fig} (bottom). Physically they all correspond
to full relativistic propagators, even though $A$ and $\sigma$ are
spatial rather than space-time tensors.

\begin{figure}[t!]
\centering \noindent
\includegraphics[width=15cm]{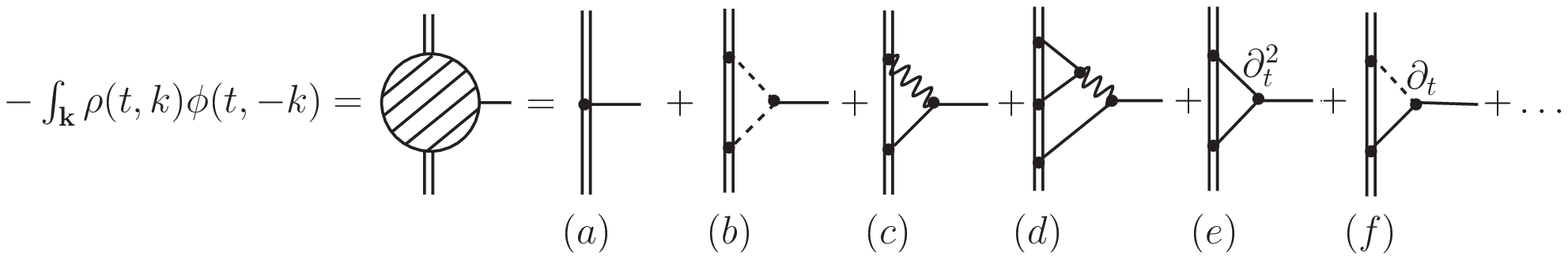}
\caption[]{Feynman diagrams which contribute to the dressed energy
distribution up to 2PN. Diagram (a) is leading (order 0PN) while
the rest are 2PN.} \label{energy-renorm-fig}
\end{figure}

\begin{figure}[t!]
\centering \noindent
\includegraphics[width=14cm]{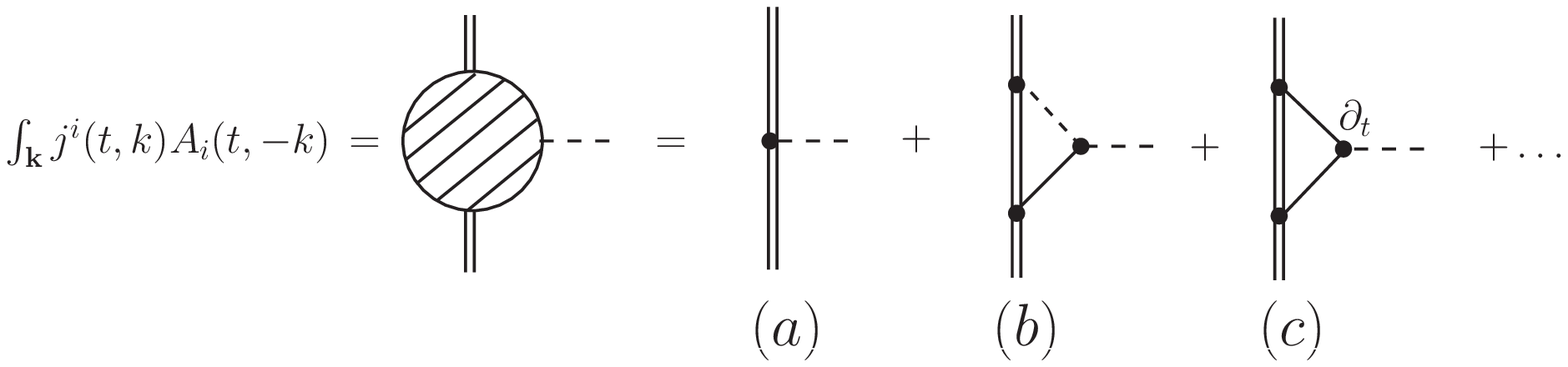}
\caption[]{Feynman diagrams which contribute to the dressed
momentum distribution up to 1.5PN. Diagram (a) is leading (order
0.5PN) while the rest are +1PN (1.5PN).}
\label{momentum-renorm-fig}
\end{figure}

\begin{figure}[t!]
\centering \noindent
\includegraphics[width=15cm]{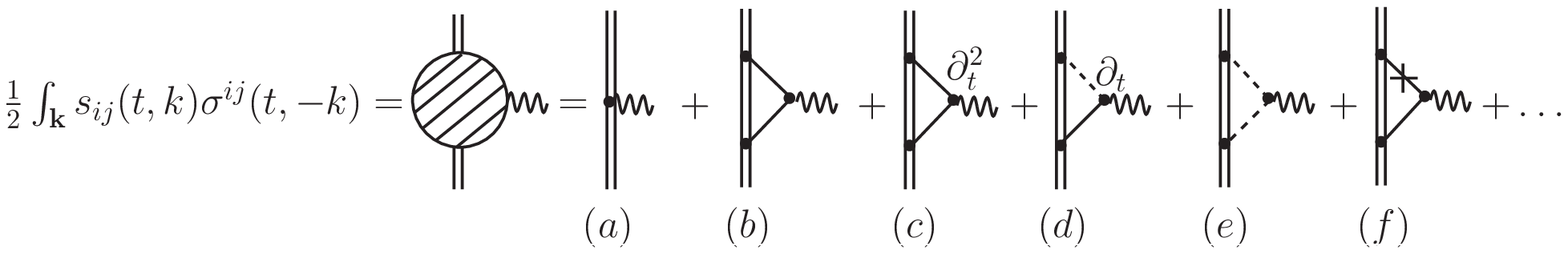}
\caption[]{Feynman diagrams which contribute to the dressed stress
tensor distribution up to 2PN. Diagrams (a,b) are leading (order
1PN) while the rest are +1PN (2PN).} \label{stress-renorm-fig}
\end{figure}

The Feynman diagrams required for the computation of the dressed
energy distribution, $\rho$, \footnote{In this section we shall
shorten the notation and write $\rho$ rather than $\rho_{dr}$.}
and the stress, $s^{ij}$, up to 2PN as well as the dressed
momentum distribution, $j^i$, up to 1.5PN are shown in figure
\ref{energy-renorm-fig}, figure \ref{momentum-renorm-fig} and
figure \ref{stress-renorm-fig}.\footnote{Alternatively, one may
count PN orders relative to the
 leading order. With this convention the dressed energy is computed to order +2PN relative to leading, while the dressed momentum and stress
 are computing to +1PN beyond leading.} The PN order was chosen as the one required for the 2PN effective action, apart for the stress where we chose to compute an additional PN order. All the results below apart from the last term of the dressed stress 2PN were tested and confirmed against the known expression for 2PN effective action. In addition we shall show in the next subsections how to use the dressed charges to calculate diagrams beyond 2PN.

Given the Feynman rules of the previous subsection,
one can write down the expressions for the Feynman integrals.
The integrals which are essential for the evaluation of the loop
integrals and the Fourier transforms are listed in Appendix
\ref{master-int}. Note also that since we compute the dressed
sources up to a definite order in the PN expansion there is no need
to keep the vertices of figure \ref{PN-Feynman-rules-1} as they are.
Instead, one has to expand each such vertex keeping only those
powers of ``$v$" which contribute to the PN order under
consideration.

We proceed to present the results of all the diagrams. We start
from the dressed energy distribution. Let us denote
 \bea
 \lambda&:=&Gm^2{\Gamma \left( {3-d \over 2} \right) \Gamma \left( {d-1 \over 2} \right) \over
 \left( 16\pi \right)^{d-4 \over 2} \Gamma \left( {d \over 2} \right)} \, ,
 \eea
then
 \bea
 \text{fig.\ref{energy-renorm-fig}}(a) &=& - m \int dt\, \frac{(d-3) + v^2}{ (d-3)\sqrt{1-v^2} }
 \, \phi(t,\vec{x}(t))
 \non
 &\Rightarrow& \delta\rho  (t, \vec{x}) = m \left( 1+ \frac{d-1}{2(d-3)}v^2 + \frac{3d-5}{8(d-3)}v^4 \right) \,
 \delta (\vec{x} - \vec {x}(t)) \, ,
 \non
 &&\delta\rho  (t, \vec{k}) = m \left( 1+ \frac{d-1}{2(d-3)}v^2 + \frac{3d-5}{8(d-3)}v^4 \right)
 \exp (-i \vec{k} \cdot \vec {x}(t)) \, ,
 \eea
 \be
 \text{fig.\ref{energy-renorm-fig}}(b) = \lambda \frac{(d-2)^2}{2(d-3)}
 \int dt \,  \vec{v}^{\, 2} \int_{\textbf{k}} \, e^{-i \vec{k} \cdot \vec{x}(t)}
 \, |\vec{k}|^{d-3} \phi(t,-\vec{k})
 \ee
 \be
 \text{fig.\ref{energy-renorm-fig}}(c) = \lambda \, {d-2 \over 2}
 \int dt \,  \int_{\textbf{k}} \, e^{-i \vec{k} \cdot \vec{x}(t)}
 \, |\vec{k}|^{d-5} (\vec{k}\cdot \vec{v})^2 \phi(t,-\vec{k})
 \ee
 \bea
 \text{fig.\ref{energy-renorm-fig}}(d) &=& { (16\pi G)^2 m^3 \over 24}
 \, { (d-3)^2 \over (d-2)(d-5)}
 \int dt \,  \int_{\textbf{k}} \, e^{-i \vec{k} \cdot \vec{x}(t)}
 \, \vec{k}^{\,2}  \, I(|\vec{k}|)\, \phi(t,-\vec{k})
 \non
 &\Rightarrow& \delta\rho  (t, \vec{k}) = -
 { (16\pi G)^2 m^3 \over 24}
 \, { (d-3)^2 \over (d-2)(d-5)}
 \, e^{-i \vec{k} \cdot \vec{x}(t)} \, \vec{k}^{\,2}  \, I(|\vec{k}|) \, ,
 \label{sing-expr}
 \eea
with
 \be
 I(|\vec{k}|)=\sqrt{\pi}{\Gamma(4-d)\over (4\pi)^{d-1}} {\Gamma(d/2-3/2)^2\Gamma(d-3)\over \Gamma(d/2-1)\Gamma(3d/2 - 9/2)}
 \({\vec k^{\,\,2} \over 2}\)^{d-4} \, .
 \ee

Apparently, the above expression possesses a pole when $d=4$, and
thus one needs to introduce a counter-term which inevitably leads
to a logarithmic behavior of the dressed energy distribution with
scale. Yet this pole unphysical. Indeed, applying a Fourier
transform to equation (\ref{sing-expr}) yields
 \be
 \delta\rho  (t, \vec{r}) =  (8 \pi G)^2\left( { m \over \Omega_{d-2} } \right)^3
 {(d-3) \over  (d-2)(d-5)} \, |\vec{r}|^{7-3d} \, ,
 \label{reg-expr}
 \ee
where
 \be
 \Omega_{d-2}={ 2 \pi^{ d-1 \over 2 } \over \Gamma \left( {d-1 \over 2} \right)} ~~ ,
 ~~ \vec{r}=\vec{x}-\vec{x}(t) \, .
 \ee
and this expression is regular for $d=4$.

Alternatively, one could stay within the wave-number space ($k$-space) by
introducing the following counter-term
 \be
 \texttt{c.t.}=  - {G^2 m^3 \over 6(d-4)}
 \int dt \,  \int_{\textbf{k}} \, e^{-i \vec{k} \cdot \vec{x}(t)}
 \, \vec{k}^{\,2} \, \phi(t,-\vec{k})
 = {G^2 m^3 \over 6 (d-4)} \int dt \, \square \phi(t,\vec{x}(t)) \, .
 \ee

However, as explained in \cite{GoldbergerRothstein1} such a term
can be removed from the Lagrangian by an appropriate field
redefinition as it is proportional to the leading order equation of
motion for the NRG field $\phi$ and hence it is a redundant term.\footnote{In quantum field theory such
an objects is called a redundant \emph{operator} referring to its action on the Hilbert space of states. In the classical theory it is not an operator and calling it a \emph{term} is more appropriate.}

The contribution to the energy distribution of the diagrams which
contain time derivatives is given by
 \bea
 \text{fig.\ref{energy-renorm-fig}}(e) &=& - {\lambda \over 8} \int dt \int_{\textbf{k}}
 \, e^{-i \vec{k} \cdot \vec{x}(t)}
 \non
 && \times \,
 \, |\vec{k}|^{d-5} \left[ \vec{k}^{\,2} \vec{v}^{\,2} - (3d-5)(\vec{k}\cdot\vec{v})^2 - 4(d-2)(i\vec{k}\cdot\vec{a} )\right]
 \, \phi(t,-\vec{k})
 \non
 \text{fig.\ref{energy-renorm-fig}}(f) &=& -{ d-2 \over 2 } \lambda \int dt
 \int_{\textbf{k}}
 \, e^{-i \vec{k} \cdot \vec{x}(t)} |\vec{k}|^{d-5} \left[ i\vec{k}\cdot\vec{a} + 2(\vec{k}\cdot\vec{v})^2 \right]
 \, \phi(t,-\vec{k})
 \eea
Combining altogether we get up to 2PN
 \begin{multline}
 \rho(t,\vec{k})\, e^{i \vec{k} \cdot \vec{x}(t)} = m \left( 1+ \frac{d-1}{2(d-3)}v^2 + \frac{3d-5}{8(d-3)}v^4 \right)
 -{ (16\pi G)^2 m^3 \over 24} \, { (d-3)^2 \over (d-2)(d-5)} \, \vec{k}^{\,2}  \, I(|\vec{k}|)
 \\- \frac{4d^2-17d+19}{8(d-3)}\lambda |\vec{k}|^{d-3}v^2+{d-3\over 8}\lambda |\vec{k}|^{d-5}(\vec{k}\cdot\vec{v})^2  \, .
 \end{multline}
 where the exponential on the left hand side really belongs on the right hand side and was moved from there to achieve a ``cleaner'' form.
 Transforming back to coordinate space yields
 \begin{multline}
 \rho(t,\vec{r})= m \left( 1+ \frac{d-1}{2(d-3)}v^2 + \frac{3d-5}{8(d-3)}v^4 \right)\delta(\vec{r})
 \\ + 8 \pi G \left( { m \over \Omega_{d-2} } \right)^2
 \[ (\vec{v}\cdot \widehat{r})^2 - 2{d-2 \over d-3} v^2\] \, |\vec{r}|^{4-2d}
 \\+(8 \pi G)^2\left( { m \over \Omega_{d-2} } \right)^3 {(d-3) \over  (d-2)(d-5)} \, |\vec{r}|^{7-3d} \, .
 \end{multline}

The results for the dressed momentum distribution up to 1.5PN are
\bea
 \text{fig.\ref{momentum-renorm-fig}}(a) &=&   \int dt\, \frac{m\,v^i }{ \sqrt{1-v^2} }
 \, A_i(t,\vec{x}(t))
 \non
 &\Rightarrow& \delta j^i  (t, \vec{x}) =  m\,v^i\(1+{1\over 2} v^2 \) \,
 \delta (\vec{x} - \vec {x}(t)) \, ,
 \non
 &&\delta j^i (t, \vec{k}) = m\,v^i\(1+{1\over 2} v^2 \)
 \exp (-i \vec{k} \cdot \vec {x}(t)) \, ,
 \non
 \text{fig.\ref{momentum-renorm-fig}}(b) &=& - \lambda \, {d-2 \over 2}
 \int dt \,  \int_{\textbf{k}} \, e^{-i \vec{k} \cdot \vec{x}(t)}
 \, |\vec{k}|^{d-3} \vec{v} \cdot \vec{A}(t,-\vec{k})
 \non
 \text{fig.\ref{momentum-renorm-fig}}(c) &=& \lambda  {d-3 \over 8(d-2)}
 \int dt   \int_{\textbf{k}} e^{-i \vec{k} \cdot \vec{x}(t)}
 |\vec{k}|^{d-5} (\vec{k}^{\,2} \, v_{i} + (d-3)(\vec{k} \cdot \vec{v})  k_i) A^i(t,-\vec{k})
 \non
 \eea

Altogether we obtain up to 1.5PN
 \begin{multline}
 j^i(t,\vec{k})\, e^{i \vec{k} \cdot \vec{x}(t)}=mv^i \(1+{1\over 2} v^2 \)
 - \, {4 d^2-17d+19 \over 8(d-2)} \, \lambda|\vec{k}|^{d-3} v^i
 \\ + \, {(d-3)^2 \over 8(d-2)} \, \lambda \, |\vec{k}|^{d-5} \, (\vec{k} \cdot \vec{v}) \, k^i \, .
 \label{momentum-charge}
 \end{multline}
In coordinate space the dressed momentum distribution is
\begin{multline}
 j^i(t,\vec{r})=mv^i \(1+{1\over 2} v^2 \)\delta(\vec{r})
  + 8 \pi G \, \left( { m \over \Omega_{d-2} } \right)^2
 \[ {d-3 \over d-2}(\vec{v}\cdot \widehat{r}) \, \widehat{r}^{~i} - 2 \, v^i\] \,
 |\vec{r}|^{\, 4-2d} \, .
\end{multline}

Finally, the results for the dressed stress charge up to 2PN are
given by
 \bea
 \text{fig.\ref{stress-renorm-fig}}(a) &=& {m \over 2} \int dt\,  { v^i\, v^j \over \sqrt{1-v^2}} \, \sigma_{ij}(t, \vec{x})
 \non
 &\Rightarrow& \delta s^{ij}(t, \vec{x}) = m \, v^i\, v^j\(1+{1\over 2} v^2 \) \,
 \delta (\vec{x} - \vec {x}(t)) \, ,
 \non
 && \delta s^{ij}(t, \vec{k}) = m \, v^i\, v^j\(1+{1\over 2} v^2 \) \,
 \exp (-i \vec{k} \cdot \vec {x}(t)) \, ,
 \eea
 \bea
 \text{fig.\ref{stress-renorm-fig}}(b) &=& \lambda \frac{(d-3)^2}{16(d-2)}
 \int dt \( 1+\frac{d-1}{d-3} \, v^2 \) \,  \int_\textbf{k} \, e^{-i \vec{k} \cdot \vec{x}(t)}
 \, |\vec{k}|^{d-5} \left[ k_i k_j - \vec{k}^{\,2} \delta_{ij} \right]
 \sigma^{ij}(t,-\vec{k})
 \non
 \eea
 \bea
 \text{fig.\ref{stress-renorm-fig}}(c) &=& \lambda
 \frac{d-3}{32(d-2)}\int dt \,  \int_\textbf{k} \, e^{-i \vec{k} \cdot \vec{x}(t)}
 \, |\vec{k}|^{d-5}
 \non
 &&\times \left[ \frac{(3d-5)^2}{d-3}(\vec{k}\cdot \vec{v})^2 + \vec{k}^{\,2}\vec{v}^{\,2}
 + { 8(d-1)(d-2) \over d-3 } i \vec{k} \cdot \vec{a} \right]
 \delta_{ij} \sigma^{ij}(t,-\vec{k})
 \eea
 \bea
 \text{fig.\ref{stress-renorm-fig}}(d) &=& \lambda {d-2 \over 2}
 \int dt \,  \int_\textbf{k} \, e^{-i \vec{k} \cdot \vec{x}(t)}
 \, |\vec{k}|^{d-5}
  \left[ \frac{1}{2} (\vec{k}\cdot \vec{v})^2 \delta_{ij} + (i \vec{k} \cdot \vec{a}) \delta_{ij}
  - i k_i a_j  \right] \sigma^{ij}(t,-\vec{k})
 \non
 \eea
 \bea
 \text{fig.\ref{stress-renorm-fig}}(e) &=& {\lambda \over 8}
 \int dt \,  \int_\textbf{k}  e^{-i \vec{k} \cdot \vec{x}(t)} |\vec{k}|^{d-5}
 \left[ (d-3) (\vec{k}^{\,2} \vec{v}^{\,2} \delta_{ij}-\vec{v}^{\,2} k_i k_j) -2(d-2) \vec{k}^{\,2} v_i v_j
  \right] \sigma^{ij}(t,-\vec{k})
 \non
 \eea
 \bea
 \text{fig.\ref{stress-renorm-fig}}(f) &=& - { \lambda \over 4}
 \frac{d-3}{d-2} \int dt \,  \int_\textbf{k} \, e^{-i \vec{k} \cdot \vec{x}(t)}
 \, |\vec{k}|^{d-5}
 \non
 &&\times \left[ \frac{d-2}{8} \vec{k}^{\,2} \vec{v}^{\,2}
 -\frac{(d-3)(d-4)}{8}
 (\vec{k} \cdot \vec{v})^2-\frac{d^2-7d+11}{4}(i\vec{k} \cdot \vec{a}) \right] \delta_{ij}\sigma^{ij}(t,-\vec{k})
 \non
 && - { \lambda \over 4}
 \frac{d-3}{d-2} \int dt \,  \int_\textbf{k} \, e^{-i \vec{k} \cdot \vec{x}(t)}
 \, |\vec{k}|^{d-5}
 \non
 &&\times \left[ - \frac{d-3}{8} \vec{v}^{\,2} k_i k_j
 +\frac{(d-3)(d-5)}{8}(\widehat{k} \cdot \vec{v})^2 k_i k_j - \frac{1}{4}\vec{k}^{\,2} v_iv_j \right] \sigma^{ij}
 \non
 &&- { \lambda \over 4}
 \frac{d-3}{d-2} \int dt \,  \int_\textbf{k} \, e^{-i \vec{k} \cdot \vec{x}(t)}
 \, |\vec{k}|^{d-5}
  \left[-\frac{ik_i a_j}{2}+\frac{(d-3)(d-5)}{4}(\vec{k} \cdot \vec{a})
 \, i\widehat{k}_i \widehat{k}_j \right] \sigma^{ij}
 \non
 \eea
As a result, the dressed stress charge up to 2PN is given by
 \bea
 s^{ij}(t,\vec{k})\, e^{i \vec{k} \cdot \vec{x}(t)} &=& m \, v^i\, v^j\(1+{1\over 2} v^2 \)
 - \, {4 d^2-17d+19 \over 4(d-2)} \, \( ik^{(i}a^{j)}+{\vec{k}^{\,2}\over 2}v^iv^j \) |\vec{k}|^{d-5}\lambda
 \non
 &+& \, {(d-3)^2 \over 8(d-2)}  \, \( 1-{v^2\over 2}-{d-5 \over2} (\vec{v} \cdot \widehat{k})^2
 -(d-5) \, {i\vec{a} \cdot \vec{k} \over \vec{k}^{\,2}}\)k^ik^j |\vec{k}|^{d-5} \lambda
 \non
 &+& \[ {(d-1)(d^2+8d-21) \over 16(d-2)}(\vec{k} \cdot \vec{v})^2 + {d^3+2d^2-12d+7 \over 8(d-2)} \, i\vec{a} \cdot
 \vec{k}\] |\vec{k}|^{d-5}\lambda \delta^{ij}
 \non
 &-&{(d-3)^2 \over 8(d-2)} \, \( 1-{v^2\over 2} \) |\vec{k}|^{d-3}\lambda \, \delta^{ij} \, ,
 \label{2PN-stress-charge}
 \eea
where $(ij)$ denotes symmetrization with respect to indices $i$ and
$j$ with factor $1/2$ included. In the coordinate space we obtain
\begin{multline}
 s^{ij}(t,\vec{r}) = m \, v^i\, v^j\(1+{1\over 2} v^2 \)\delta(\vec{r})-{(3d-5) \over (d-2)(d-3)} f(r) \vec{r}^{\,(i}a^{j)}-2f(r)v^iv^j
 \\-{d^2+10d-15 \over 4(d-2)(d-3)}f(r)
 \(v^2-2(d-2)(\vec{v}\cdot \widehat{r})^2\)\delta^{ij}
 \\-{d-3 \over d-2} f(r)\, \( (d-1)(\vec{v}\cdot \widehat{r})^2\widehat{r}^{\,i}\widehat{r}^{\,j}
 -2(\vec{v}\cdot \widehat{r})v^i\widehat{r}^{\,j} +
 \frac{1}{2}\( 1-\frac{v^2}{2} \)\delta^{ij}-\widehat{r}^{\,i}\widehat{r}^{\,j} +(\vec{a}\cdot \vec{r})\widehat{r}^{\,i}\widehat{r}^{\,j}\)
 \\+{d^2+5d-8 \over 2(d-2)(d-3)}f(r)(\vec{a}\cdot \vec{r}) \, \delta^{ij} \, .
\end{multline}
with
 \be
 f(r)={8 \pi G \over |\vec{r}|^{\, 2d-4} } \left( { m \over \Omega_{d-2} } \right)^2
 ~.
 \ee

\subsection{Skeletons for 2PN and 3PN}
\label{PNskeletons-subsection}

We shall now demonstrate how all the bare diagrams of the two-body effective action up to 2PN, order by order transform into their dressed form, including their corresponding skeletons.

{\bf At 0PN} (Newtonian order) a single diagram (figure \ref{0PN-fig}) contributes representing the interaction of two tree-level masses.

\begin{figure}[t!]
\centering \noindent
\includegraphics[width=9cm]{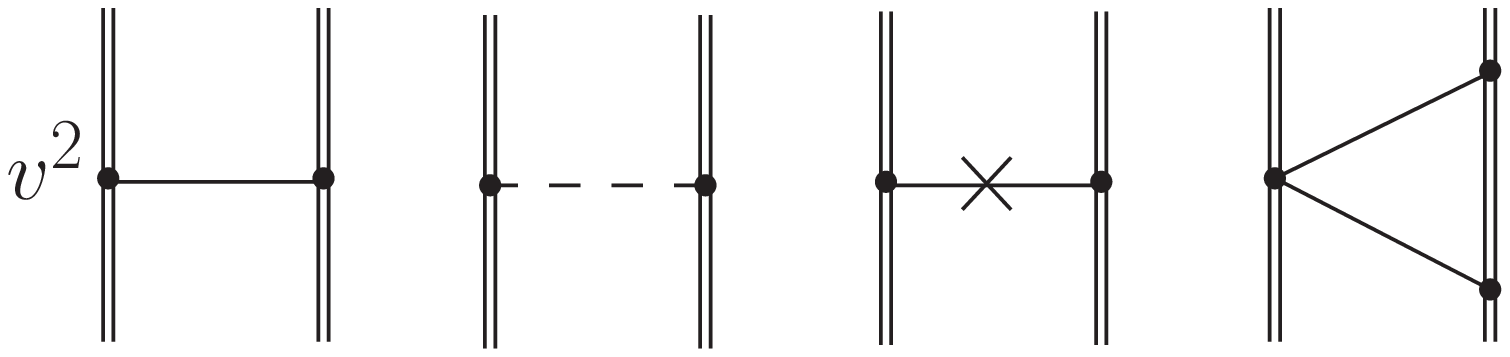}
\caption[]{The four diagrams which contribute to the two-body
effective action at 1PN -- The Einstein-Infeld-Hoffmann
Lagrangian.} \label{1PN-fig}
\end{figure}

{\bf At 1PN} (Einstein-Infeld-Hoffmann Lagrangian) the four
diagrams shown in figure \ref{1PN-fig} contribute. The first is a
$v^2$ component of the mass vertex and can be considered to be a
trivial dressing of the energy. The second is the tree-level
interaction of two currents. The third represents a propagator
dressing (retardation) of the Newtonian potential. Finally the
fourth is due to a non-linear world-line vertex which accounts for
the gravitating nature of potential energy. Altogether none is
irreducible with a topology other than Newtonian. We note that the
last diagram is of order $\co\(G^2 m^3 v^0\)$ where $m$ represents
a typical mass and $v$ a typical velocity while all the previous
diagrams are of order $\co\(G m^2 v^2\)$.

{\bf At 2PN} Gilmore and Ross \cite{GilmoreRoss} found 21 diagrams. We
shall see that many can be interpreted to represent
dressing effects while only one is both non-factorizable
and dressing-irreducible.
Indeed \cite{GilmoreRoss} find its computation to be the core or
essential computation at this order.

\begin{figure}[t!]
\centering \noindent
\includegraphics[width=11cm]{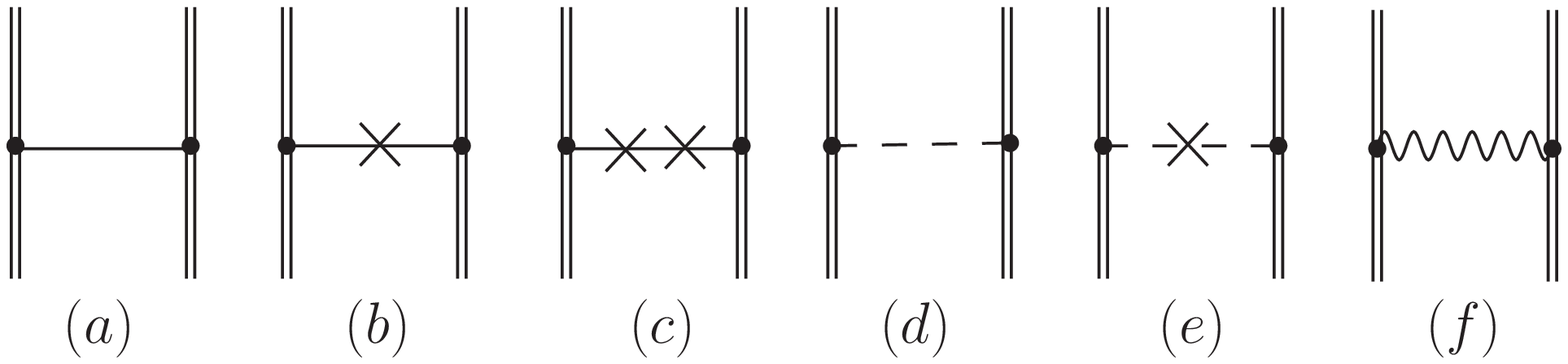}
\caption[]{2PN Diagrams contributing at order $\co\(G m^2 v^4\)$
(following \cite{GilmoreRoss}).} \label{2pnG1-fig}
\end{figure}

The six diagrams (a-f) shown in figure \ref{2pnG1-fig} contribute at
order $\co\(G m^2 v^4\)$. Diagrams (b,c,e) represent propagator
dressing to diagrams from lower orders,  while the rest
involve only tree-level $v$-dependent vertices.

\begin{figure}[t!]
\centering \noindent
\includegraphics[width=11cm]{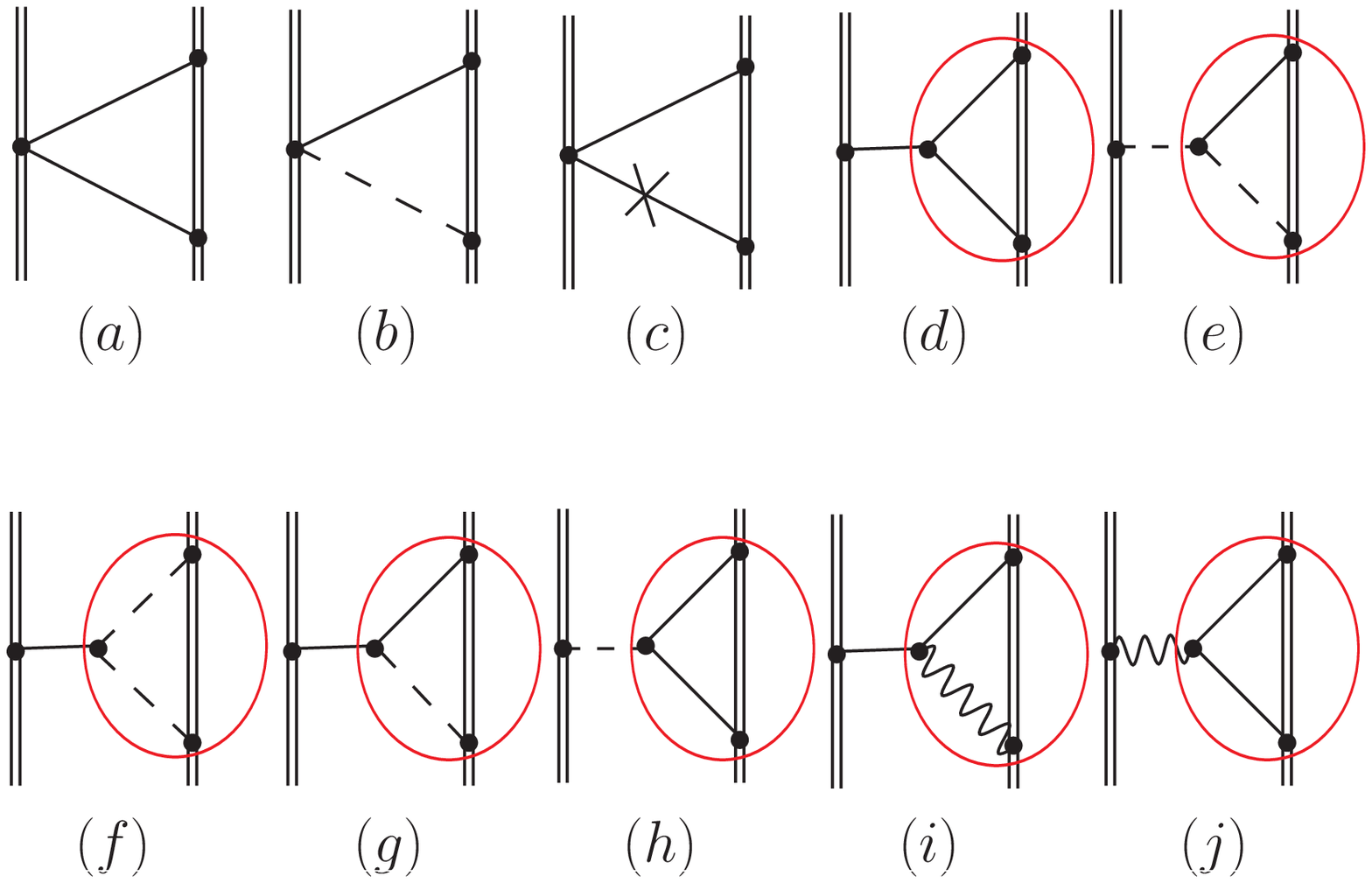}
\caption[]{2PN Diagrams contributing at order $\co\(G^2 m^3 v^2\)$
(following \cite{GilmoreRoss}).} \label{2pnG2-fig}
\end{figure}

At order $\co\(G^2 m^3 v^2\)$ there are 10 diagrams shown in figure \ref{2pnG2-fig}. Diagrams (a,b,c) are V-shaped and as such
factorizable and at least one factor is the Newtonian potential.
Diagrams (d-j) are Y-shaped and as such are dressing-reducible. Four
diagrams (d,f,g,i) represent mass dressing, the two (e,h)
represent current dressing, while finally (j) can be thought
to represent stress dressing.

At order $\co\(G^3 m^4 v^0\)$ there are 5 diagrams. Both (a) and (c)
factorize into 3 Newtonian-potential factors. Diagrams (b)
represents a mass dressing (the circled piece) while (e)
includes two $\sigma$ dressing sub-diagrams. Finally diagram
(d) is the one and only truly irreducible diagram at 2PN. Actually
for some yet-unexplained reason the computation reduces after
several steps to a square of the master one-loop integral. We
speculate that this is special to the GR action (and is not generic
to classical field theories) and in particular to the gauge symmetry
which may relate this diagram to other diagrams at 2PN.

\begin{figure}[t!]
\centering \noindent
\includegraphics[width=12cm]{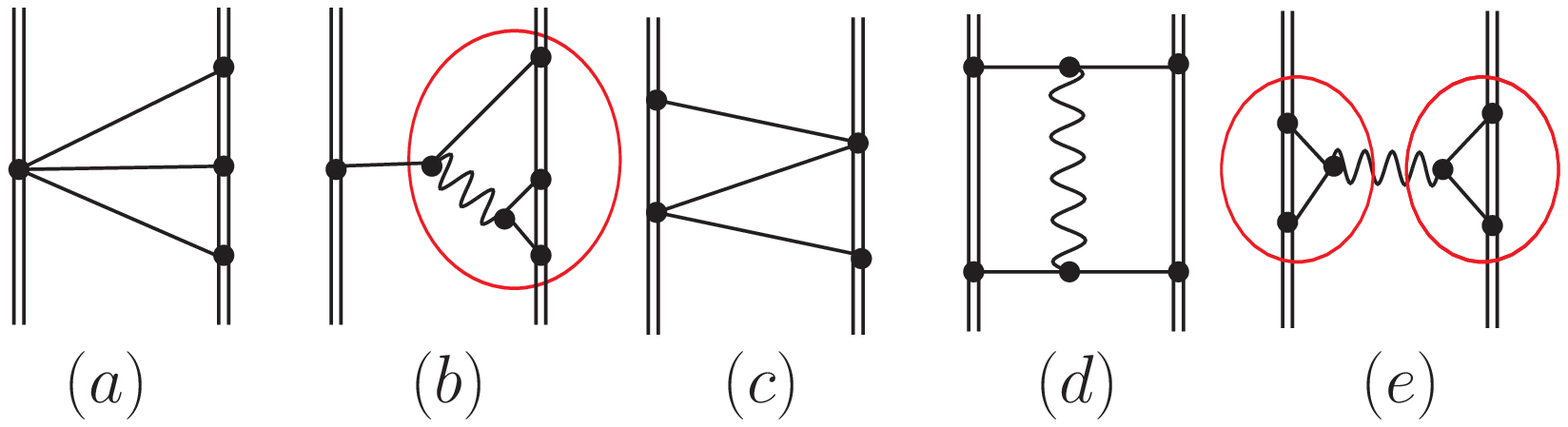}
\caption[]{2PN Diagrams contributing at order $\co\(G^3 m^4 v^0\)$
(following \cite{GilmoreRoss}).} \label{2pnG3-fig}
\end{figure}

\begin{figure}[t!]
\centering \noindent
\includegraphics[width=12cm]{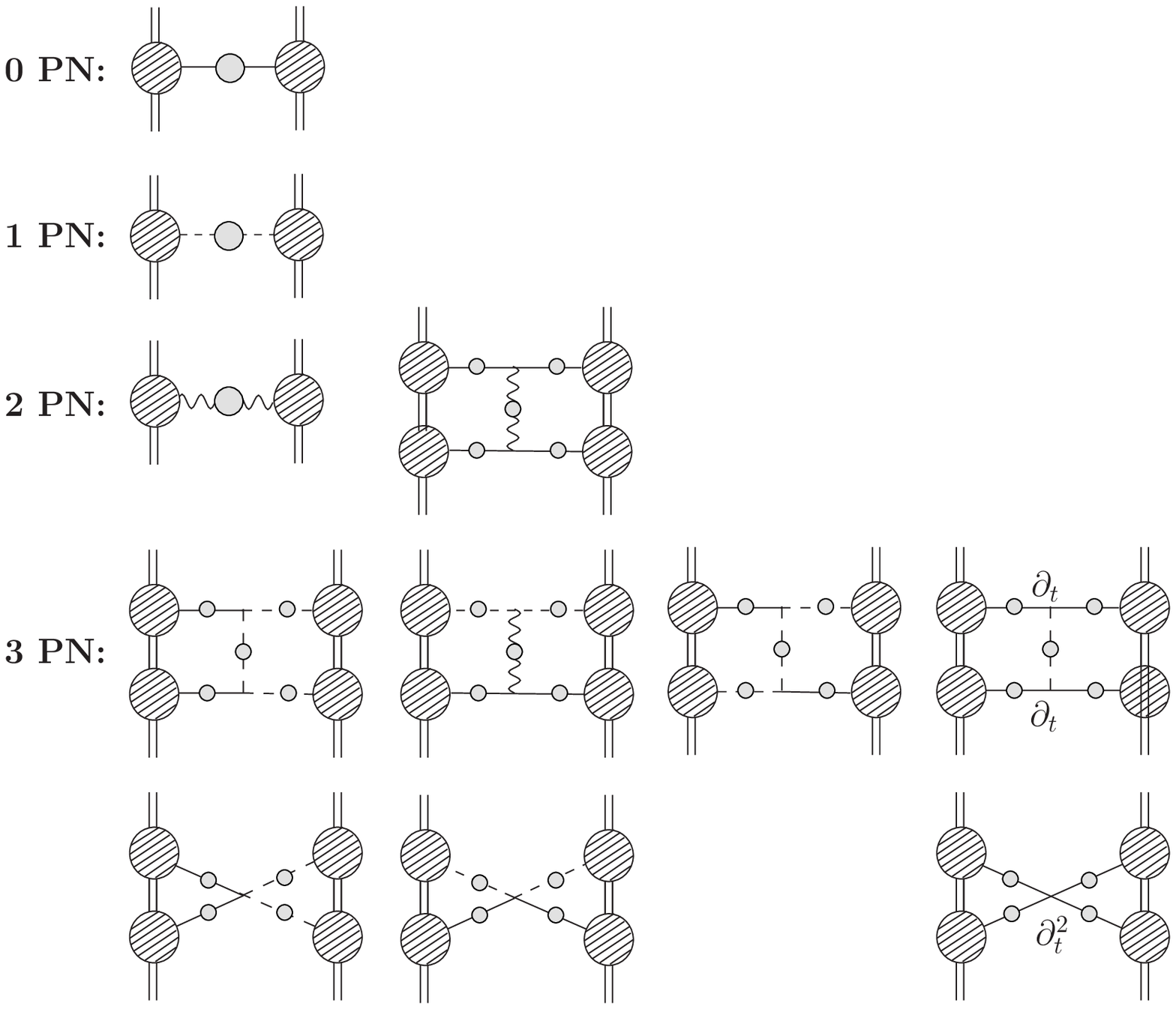}
\caption[]{A listing of all non-factorizable skeletons appearing
in the dressed Post-Newtonian perturbation theory up to 3PN,
 listed by the PN order at which they first appear.} \label{PN-skeletons-fig}
\end{figure}

As a result of this analysis we can extract all the non-factorizable skeletons up to 2PN.
 These are listed in figure \ref{PN-skeletons-fig}, where the skeletons are labeled according to the PN order in which they first appear.
By evaluating the dressed diagrams we successfully tested our expressions for the dressed charges from the previous subsection against the known
 expressions for the effective action.

{\bf 3PN}. As a step towards the determination of 3PN we extend the list of skeletons up to that order using the classification of possible topologies in figure \ref{irred-top-fig} and some knowledge on the Feynman rules of PN. Taken together with the evaluation of the dressed vertices (partially obtained in the last subsection) this list of skeletons leads the way to a determination of the 3PN part of the 2-body effective action. We note that the 3-loop topologies of figure \ref{irred-top-fig} are \emph{not} realized at 3PN. The reason is the absence in PN of certain bulk vertices such as $\phi^3$.

\subsection{Computing beyond 2PN}
\label{computing-beyond-subsection}

In this subsection we explicitly demonstrate the economizing
ability of proposed approach in several cases \bi \item A certain
economization during 2PN calculation. \item Using the dressed
energy distribution computed above in subsection
\ref{PNdressed-charges-subsection} to calculate terms of the
Newtonian interaction type of order 3PN and 4PN. \item Similarly,
using the computed dressed momentum distributions we calculate a
current-current interaction term of order 3PN. \item Finally, we
use our highest order results for the dressed stress in order to
compute 3PN terms by attaching the external $\sigma$ leg to the
second compact object.
 \ei

\begin{figure}[t!]
\centering \noindent
\includegraphics[width=9cm]{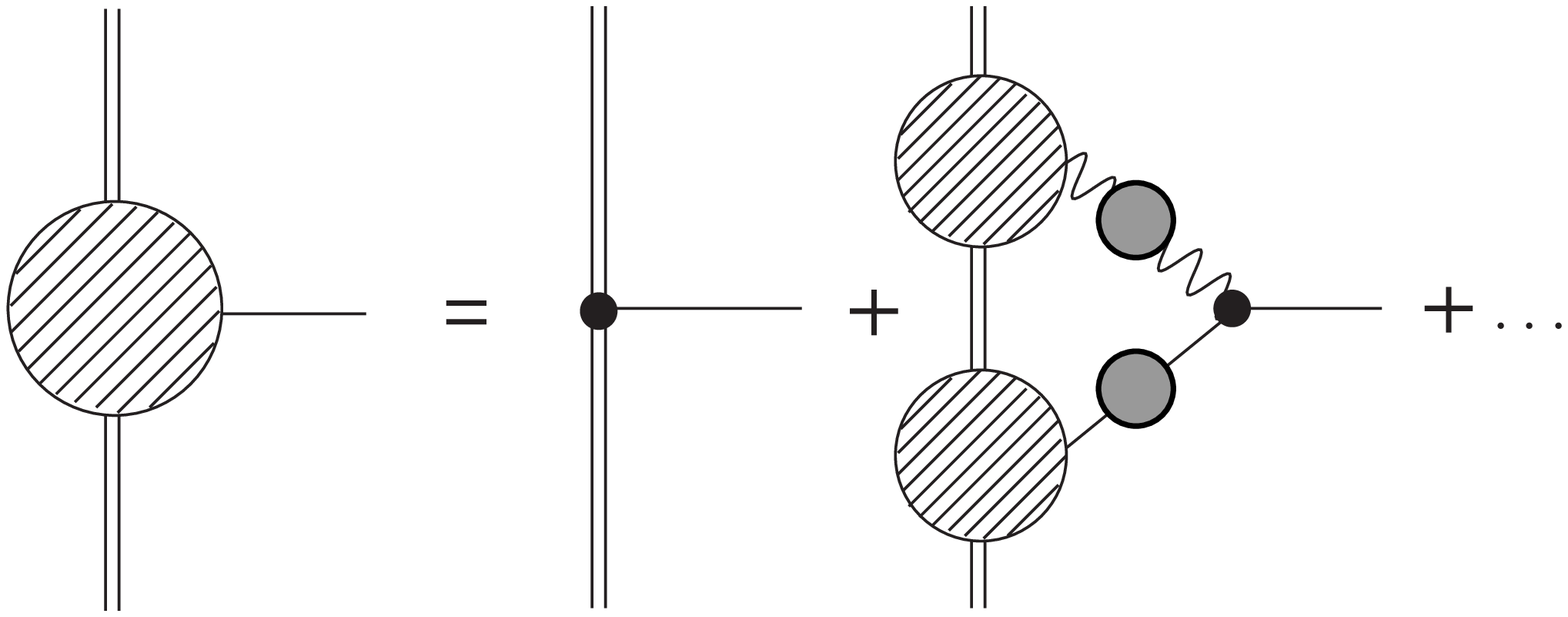}
\caption[]{The diagrammatic representation of the recursive integral
equation satisfied by the dressed energy.} \label{PN-recursion-fig}
\end{figure}

{\bf An application at 2PN}. The first non-trivial example appears during the computation
of the 2-body effective Lagrangian at order 2PN.
Indeed the dressed stress sub-diagram in figure \ref{stress-renorm-fig}(b) appears twice at 2PN --
both at figure \ref{2pnG2-fig}(j) and at figure \ref{2pnG3-fig}(b), and of course it is enough to evaluate it once.
Another point of view is to obtain both dressed energies corresponding to figures \ref{energy-renorm-fig}(c,d) from the recursive relation shown in
figure \ref{PN-recursion-fig} after substituting in the leading contribution to the dressed stress from figures \ref{stress-renorm-fig}(a,b).

\begin{figure}[t!]
\centering \noindent
\includegraphics[width=12cm]{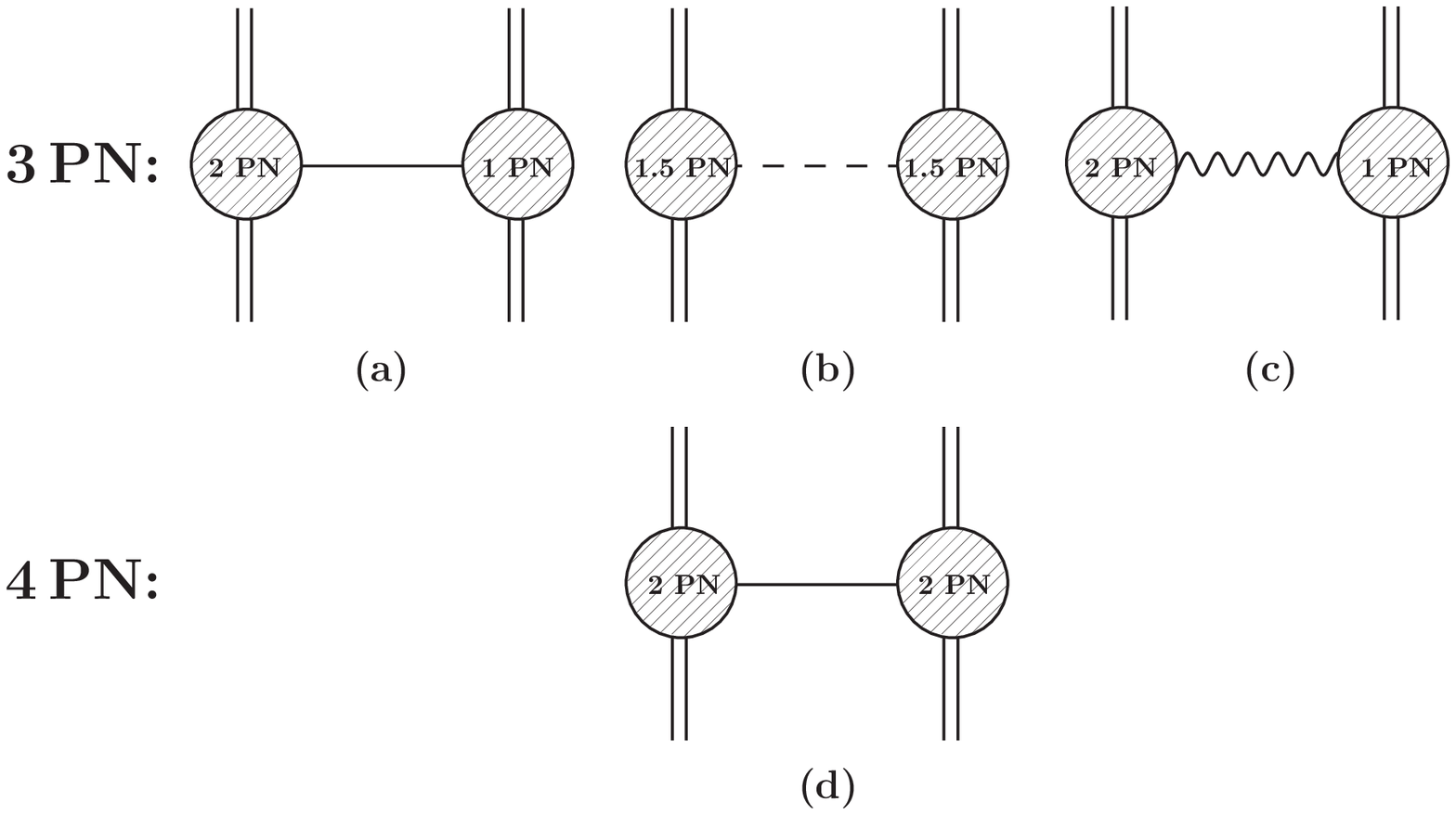}
\caption[]{The diagrammatic representation of terms beyond 2PN in the 2-body effective action: (a) a 3PN $\rho$-$\rho$ term, (b) a 3PN $j$-$j$ term, (c) a 3PN $s$-$s$ term and finally (d) a 4PN $\rho$-$\rho$ term.}
 \label{3-4PN-fig}
\end{figure}

Proceeding {\bf beyond 2PN} we shall now compute certain 3PN and 4PN terms
in the 2-body effective action building on the results of the
previous section. Pictorially these terms are shown on
figure \ref{3-4PN-fig}, where the bubbles represent the dressed
charges, and we explicitly indicate on each bubble which piece of
the dressed charge is essential to the computation.

 Building on the Feynman rules of the previous section, we obtain
  \bea
 \text{fig.}\ref{3-4PN-fig}(a,d) &=& 8 \pi G \, \frac{d-3}{d-2}
 \int dt \int_{\textbf{k}}{\rho_2( \,t,-\vec{k}\,)\rho_1(\,t,\vec{k}\,) \over \vec{k}^{\,2}}
 \, ,
 \label{3-4PN-scalar-expr}\\
 \text{fig.}\ref{3-4PN-fig}(b) &=& - 16 \pi G \,\int dt \int_{\textbf{k}}{\vec{j}_2( \,t,-\vec{k}\,) \cdot \vec{j}_1(\,t,\vec{k}\,) \over \vec{k}^{\,2}}
 \, ,
 \label{3PN-vector-expr}\\
 \text{fig.}\ref{3-4PN-fig}(c) &=& 8 \pi G
 \int dt \int_{\textbf{k}} {s_2^{\,ij}( \,t,-\vec{k}\,)\,P_{ij,kl} \,s_1^{\,kl}(\,t,\vec{k}\,) \over \vec{k}^{\,2}}
 \, ,
 \label{3PN-tensor-expr}
 \eea

{\bf The 3PN $j$-$j$ and $s$-$s$ terms}. Substituting (\ref{momentum-charge}),(\ref{2PN-stress-charge}) and
$d=4$ into (\ref{3PN-vector-expr}),(\ref{3PN-tensor-expr}) yields
 \begin{multline}
 \text{fig.}\ref{3-4PN-fig}(b) = - \int dt \, {G m_1 m_2 \over
 r}(\vec{v}_1\cdot \vec{v}_2)\,v_1^{2}\,v_2^{2}
 \\
 -\int dt \, {G m_1 m_2 \over 2 r}\( {Gm_2 \over r} \, v_1^2 + {Gm_1 \over r} v_2^2\)
 \[ \, 7(\vec{v}_1\cdot \vec{v}_2) + (\vec{v}_1\cdot \widehat{r})(\vec{v}_2\cdot \widehat{r})\]
 \\
 +\frac{7\pi^2}{32}\int dt \, {G^3m_1^2m_2^2 \over r^3}
 \(\vec{v}_1\cdot \vec{v}_2-3(\vec{v}_1\cdot \widehat{r})(\vec{v}_2\cdot \widehat{r})\) \, ,
 \end{multline}

\begin{multline}
 \text{fig.}\ref{3-4PN-fig}(c)=\int dt{Gm_1m_2 \over r}
 \[ (\vec{v_1}\cdot\vec{v_2})^2 - v_1^2 v_2^2\] v_1^2+\int dt\,{G^2m_1^2\, m_2 \over r^2}
 \[ {1\over 2}(\vec{v}_1\cdot\vec{v}_2)^2+{203\over12}v_2^2v_1^2 \]
 \\+\int dt\,{G^2m_1^2\, m_2 \over r^2} \[ -{22 \over 3}(\vec{v}_2\cdot \vec{r})(\vec{a}_1\cdot \vec{v}_2)
 -{121\over 4}v_2^2(\vec{a}_1\cdot \vec{r})-{1\over6}(\vec{a}_1\cdot \vec{r})(\vec{v}_2\cdot\widehat{r})^2\]
 \\ -\int dt\,{G^2m_1^2\, m_2 \over r^2} \[ {245 \over 6}v_2^2(\vec{v}_1\cdot \widehat{r})^2
 +{1 \over 4}v_1^2(\vec{v}_2\cdot \widehat{r})^2 +{1\over 3}(\vec{v}_1\cdot \widehat{r})^2(\vec{v}_2\cdot \widehat{r})^2
 +{1 \over 12}(\vec{v}_1\cdot\vec{v}_2)^2\]
 \\+{1\over 3} \int dt\,{G^2m_1^2\, m_2 \over r^2} \[ (\vec{v_1}\cdot\vec{v_2})(\vec{v}_1\cdot \widehat{r})(\vec{v}_2\cdot \widehat{r}) \]
 +(1\leftrightarrow 2)~ .
 \\
\end{multline}
where $\vec{r}(t)=\vec{x}_2(t)-\vec{x}_1(t)$ denotes the radius
vector between the particles, and we have neglected the contact term
proportional to $\delta(\vec{r})$ since in CLEFT particles are
widely separated from each other.

{\bf The $\rho$-$\rho$ terms.} Let us now evaluate figure \ref{3-4PN-fig}(b). First note, that since
equation (\ref{sing-expr}) contains a simple pole in $d=4$ one
concludes that the integrand of (\ref{3-4PN-scalar-expr}) possesses a double
pole, and thus in contrast to the previous case (\ref{reg-expr})
a Fourier transform will not lead to a regular expression.

Therefore, we expand equation (\ref{sing-expr}) around $d=4$
 \begin{multline}
 \text{fig.\ref{energy-renorm-fig}}(d) = G^2 m_r^3 L^{-\epsilon}
 \, \int dt \,  \int_{\textbf{k}} \, e^{-i \vec{k} \cdot \vec{x}(t)}
 \, \vec{k}^{\,2} \, \phi(t,-\vec{k})
 \\ \times \[ -{1 \over 6\epsilon}+{1 \over 12} \( -1+2\gamma-2\ln(8\pi^2)+2\ln(\vec{k}^{\,2}L^2) \)+\mathcal{O}(\epsilon) \] \, ,
 \label{sing-expr2}
 \end{multline}
where $\gamma$ is Euler's constant and motivated by the QFT
renormalization approach we introduced an arbitrary length scale $L$
and the following definitions\footnote{Index "$r$" stands for
"renormalized", though in the current work we will not encounter
divergences associated with non-trivial RG flow of the mass.}
 \be
 \epsilon=4-d ~,~ G m = G m_r L^{-\epsilon} \, .
 \ee

In order to eliminate the divergence as $\epsilon \rightarrow 0$, we
add the following counter-term to the effective action
 \be
 \texttt{c.t.} = c  \int dt \, \Delta \phi(t,\vec{x}(t))
 = - c  \int dt \,  \int_{\textbf{k}} \, e^{-i \vec{k} \cdot \vec{x}(t)} \, \vec{k}^{\,2} \, \phi(t,-\vec{k}) \, ,
 \label{counterterm}
 \ee
where
 \be
 c = L^{-\epsilon} \( c_r  - { G^2 m_r^3  \over 6\epsilon} \) \, .
 \ee
Since the renormalization scale $L$ introduced above is arbitrary,
we must have
 \bea
 0&=&L{d m \over dL}=L^{-\epsilon}\(-\epsilon \, m_r + L{d  m_r \over dL}\)
 \non
 &\Rightarrow& L{dm_r \over dL} = \epsilon \, m_r \quad ,
 \non
 0&=&L{dc \over dL}=-\epsilon L^{-\epsilon} \( c_r  - { G^2 m_r^3  \over 6\epsilon} \)
 + L^{-\epsilon} \(L{dc_r \over dL}- { G^2 m_r^2  \over 2 \epsilon} ~ L{dm_r \over dL} \)
 \non
 &\Rightarrow& L{dc_r \over dL} = \epsilon \, c_r + { G^{\,2} m_r^3  \over 3 } ~ .
 \eea

Apparently, the theory exhibits a non-trivial RG flow. But as argued
in \cite{GoldbergerRothstein1} this scaling is not physical in
nature and can be removed by a suitable field redefinition which is
tantamount to a coordinate transformation. Indeed, combining
(\ref{sing-expr2}) with (\ref{counterterm}) yields
 \bea
 \delta \rho(t,\vec{k}) =
  \[  c_r - {G^2 m_r^3 \over 12} \( -1+2\gamma-2\ln(8\pi^2)+2\ln(\vec{k}^{\,2}L^2) \) \] \, e^{-i \vec{k} \cdot \vec{x}(t)} \, \vec{k}^{\,2} \, .
 \eea
In coordinate space this expression is given by
 \bea
 \delta \rho(t,\vec{r}) =
  \[  c_r - {G^2 m_r^3 \over 12} \( -1+2\gamma-2\ln(8\pi^2) \) \] \delta(\vec{r}) - {1 \over 2\pi}{G^2m_r^3 \over r^5}\, ,
 \eea
where $\vec{r}=\vec{x}-\vec{x}(t)$ and the relation
 \be
 \ln \vec{k}^{\,2} = \lim_{\al\rightarrow0}{d \over d\al}\({k}^{\,2}\)^{\al}
 \label{Log-as-power}
 \ee
was used in order to apply the  Fourier transform formulas of
Appendix \ref{master-int}.

However, the contact term in the above expression for $\delta \rho$
can be eliminated by the following field redefinition
 \be
 \phi\rightarrow \phi+4\pi G \[  c_r - {G^2 m_r^3 \over 12} \( -1+2\gamma-2\ln(8\pi^2) \) \] \delta(\vec{r}) \, .
 \label{field-shift}
 \ee
As a result, we reproduced equation (\ref{reg-expr}) for $d=4$ and
also obtained the regularized expression for $\rho(t,\vec{k})$ in
four dimensions\footnote{We omit index "$r$" to avoid abuse of
notation.}
 \begin{multline}
 \rho(t,\vec{k})\, e^{i \vec{k} \cdot \vec{x}(t)} = m \left( 1+ \frac{3}{2}v^2 + \frac{7}{8}v^4 \right)
  + \frac{\pi}{8} \, G m^2 |\vec{k}| \(15v^2-(\widehat{k}\cdot\vec{v})^2 \)
  - {G^2 m^3 \over 6} \, \vec{k}^{\,2} \ln(\vec{k}^{\,2}L^2)  \, .
 \label{2PN-mass-density}
 \end{multline}

A comment should be made regarding the field redefinition
(\ref{field-shift}). Obviously, such a shift introduces an extra
term to the world-line action as a side effect. This term is
proportional to the second derivative of the NRG field $\phi$ with
respect to time. However, it can be removed by including additional
counter-term given by
 \be
 \texttt{c.t.} = \tilde{c}  \int dt \, {\del^2\phi \over \del t^2}(t,\vec{x}(t)) \, .
 \ee
We will not elaborate the details of this elimination as they are
irrelevant to the computation of the 4PN term.

Substituting (\ref{2PN-mass-density}) into (\ref{3-4PN-scalar-expr})
leads to
\begin{multline}
\text{fig.}\ref{3-4PN-fig}(a)={21 \over 16}\int dt{Gm_1m_2 \over r}
v_1^2 v_2^4
 \\+{3\over 2}\int dt\,{G^2m_2^2\, m_1 \over r^2}v_1^2 \[ {7 \over 2}\, v_2^2 + {(\vec{v}_2\cdot \widehat r)^2 \over 2}
 +{1 \over 3}{G\,m_2 \over r}\]
 +(1\leftrightarrow 2) ~ ,
\end{multline}
and
\begin{multline}
 \text{fig.}\ref{3-4PN-fig}(d)= {49 \over 64}\int dt{Gm_1m_2 \over r} v_1^4 v_2^4
 +{7\over 8}\int dt\,{G^2m_2^2\, m_1 \over r^2}v_1^4 \[ {7 \over 2}\, v_2^2 + {(\vec{v}_2\cdot \widehat r)^2 \over 2}
 +{1 \over 3}{G\,m_2 \over r}\]
 \\ \quad\quad\quad\quad\quad\quad\quad\quad\quad\quad\quad\quad
 +{7\over 8}\int dt\,{G^2m_1^2\, m_2 \over r^2}v_2^4 \[ {7 \over 2}\, v_1^2 + {(\vec{v}_1\cdot \widehat r)^2 \over 2}
 +{1 \over 3}{G\,m_1 \over r}\]
 \\
 +{\pi^2 \over 64}\int dt\,{G^3m_1^2\, m_2^2 \over r^3} [ (\vec{v}_1\cdot \vec{v}_2)^2-{59\over 2}v_1^2v_2^2
 +{87 \over 2} v_1^2(\vec{v}_2\cdot \widehat r)^2+ {87 \over 2}v_2^2(\vec{v}_1\cdot \widehat r)^2
 \\\quad\quad\quad\quad\quad\quad\quad\quad\quad\quad\quad\quad\quad
 +{15\over 2}(\vec{v}_2\cdot \widehat r)^2(\vec{v}_1\cdot \widehat r)^2
 -6(\vec{v}_2\cdot \widehat r)(\vec{v}_1\cdot \widehat r)(\vec{v}_1\cdot \vec{v}_2)]
 \\
 +{1\over 3}\int dt\,{G^4m_1^2\, m_2^3 \over r^4}\[ v_1^2\( {23 \over 2}-8\gamma -8\ln(r/L) \)
 +(\vec{v}_1\cdot \widehat r)^2\( -{3\over 2}+2\gamma +2\ln(r/L) \)\]
 \\
 +{1\over 3}\int dt\,{G^4m_1^3\, m_2^2 \over r^4}\[ v_2^2\( {23 \over 2}-8\gamma -8\ln(r/L) \)
 +(\vec{v}_2\cdot \widehat r)^2\( -{3\over 2}+2\gamma +2\ln(r/L) \)\]
 \\ \quad\quad\quad\quad\quad\quad\quad\quad\quad\quad\quad\quad\quad\quad\quad\quad\quad\quad\quad
 -{4\over 9}\int dt\,{G^5m_1^3\, m_2^3 \over r^5}\[ 3 -2\gamma -2\ln(r/L)\] ~ .
 \
\end{multline}
To derive this expression we first applied (\ref{Log-as-power}) and
then used transform Fourier master integrals of Appendix
\ref{master-int}. Appearance of an arbitrary renormalization scale
$L$ in the above expression should not be taken as a strong evidence
for a non-trivial RG flow, since if other terms in the full 4PN
potential are considered then cancelation of logs might occur.

\subsection*{Acknowledgements}

It is a pleasure to thank Gabriele Veneziano for his lectures series ``Transplanckian scattering'' (Institute for Advanced Studies at
Jerusalem, March 2009, based in part on \cite{ACV1993,ACV2008})
which catalyzed the beginning of this work;  Ofer Aharony, Ira Rothstein and Gerhard Sch\"{a}fer for very useful comments on the manuscript;
and finally the organizers of the following meetings where part of this work was performed:
``The fifth Crete regional meeting in String Theory'' (Crete, June-July 09), and
``Gravity - New perspectives from strings and higher dimensions'' (Benasque, July 09).
 In addition MS thanks the Perimeter Institute for their kind hospitality during the completion of this work.

This research is supported by The Israel Science Foundation grant
no 607/05, by the German Israel Cooperation Project grant DIP H.52, and the Einstein Center at the Hebrew University.

\appendix
\section{Master integrals} \label{master-int}

In this appendix we present master integrals which we found
useful during the calculations presented in the text. We start with

 \be
 J=\int {d^{d} \textbf{q} \over (2\pi)^{d}} \frac{1}{ (\textbf{q}^{\,2})^{\alpha} [(\textbf{q}-\textbf{k})^{\,2}]^\beta}
 ={(\textbf{k}^{\,2})^{d/2-\al-\bt} \over (4\pi)^{d/2}} {\Gamma(\al+\bt-d/2) \over
 \Gamma(\al)\Gamma(\bt)} {\Gamma(d/2-\al)\Gamma(d/2-\bt) \over \Gamma(d-\al-\bt)}
 \label{J}
 \ee
 \begin{equation}
 J_i=\int {d^{d} \textbf{q} \over (2\pi)^{d}} \frac{q_i}{ (\textbf{q}^{\,2})^{\alpha}[(\textbf{q}-\textbf{k})^{\,2}]^\beta}=
 { d/2-\al \over d-\al-\bt} \,  J \, k_i
 \label{J_i}
 \end{equation}
 \begin{multline}
 J_{ij}=\int {d^{d} \textbf{q} \over (2\pi)^{d}} \frac{q_i \, q_j}{ (\textbf{q}^{\,2})^{\alpha} [(\textbf{q}-\textbf{k})^{\,2}]^\beta}
 \\ ={1 \over (4\pi)^{d/2}} {\Gamma(\al+\bt-d/2-1) \over
 \Gamma(\al)\Gamma(\bt)} {\Gamma(d/2-\al+1)\Gamma(d/2-\bt) \over \Gamma(d-\al-\bt+2)}
 \\  \times \[ (d/2-\al+1)(\al+\bt-d/2-1) k_ik_j + (d/2-\bt){ \textbf{k}^{\,2} \over 2}\dl_{ij} \]
 (\textbf{k}^{\,2})^{d/2-\al-\bt}
 \label{J_ij}
 \end{multline}
 \begin{multline}
 J_{ijk}=\int {d^{d} \textbf{q} \over (2\pi)^{d}}
 \frac{q_i \, q_j \, q_k}{ (\textbf{q}^{\,2})^{\al} [(\textbf{q}-\textbf{k})^{\,2}]^\bt}
 \\ ={(\textbf{k}^{\,2})^{d/2-\al-\bt} \over (4\pi)^{d/2}} {\Gamma(\al+\bt-d/2-1) \over
 \Gamma(\al)\Gamma(\bt)} {\Gamma(d/2-\al+2)\Gamma(d/2-\bt) \over \Gamma(d-\al-\bt+3)}
 \\  \times \[ (d/2-\al+2)(\al+\bt-d/2-1) k_ik_jk_k + (d/2-\bt){ \textbf{k}^{\,2} \over 2}(\dl_{ij}k_k+\dl_{jk}k_i+\dl_{ik}k_j) \]
 \label{J_ijk}
 \end{multline}
 \begin{multline}
 J_{ijkl}=\int {d^{d} \textbf{q} \over (2\pi)^{d}}
 \frac{q_i \, q_j \, q_k \, q_l}{ (\textbf{q}^{\,2})^{\al} [(\textbf{q}-\textbf{k})^{\,2}]^\bt}
 ={(\textbf{k}^{\,2})^{d/2-\al-\bt} \over (4\pi)^{d/2}} {\Gamma(d/2-\al+2)\Gamma(d/2-\bt) \over \Gamma(d-\al-\bt+4)}
 \\  \times \, {\Gamma(\al+\bt-d/2-2) \over \Gamma(\al)\Gamma(\bt)} \bigl[
 (d/2-\bt)(d/2-\bt+1)(\dl_{ij}\dl_{kl}+\dl_{ik}\dl_{jl}+\dl_{il}\dl_{jk}){(\textbf{k}^{\,2})^2 \over 4}
 \\+ \, (\dl_{ij}k_kk_l+\dl_{ik}k_jk_l+\dl_{il}k_jk_k+\dl_{jk}k_ik_l+\dl_{jl}k_ik_k+\dl_{kl}k_ik_j){\textbf{k}^{\,2}\over 2}
 \\ ~~~~~~~~~~~~~~~~~~~~~~~~~~~~~~~~~~~~~\, \times(\al+\bt-d/2-2)(d/2-\al+2)(d/2-\bt)
 \\+k_ik_jk_kk_l(\al+\bt-d/2-2)(\al+\bt-d/2-1)(d/2-\al+2)(d/2-\al+3)
 \bigr]
 \label{J_ijkl}
 \end{multline}
As noticed in \cite{GilmoreRoss} the above integrals
(\ref{J_i})-(\ref{J_ijkl}) of vector nature can be reduced to a
scalar integral (\ref{J}) on the basis of their transformation
properties under rotations. Yet such a reduction becomes involved
when the rank of the tensor under consideration increases, and therefore we choose to list all those which were relevant to this work.

In order to evaluate the above integrals we proceed as follows. We
first apply the generalized Feynman parametrization
\cite{Peskin:1995ev}
 \bea
 \prod_{i=1}^N\frac{1}{ A_i^{m_i}}=\int_0^1 \prod_{i=1}^N dx_i
 \, \delta \( \sum x_i - 1 \) \frac{\prod x_i^{m_i-1}}{\[\sum x_i A_i\]^{\sum m_i}}
 \frac{\Gamma(m_1+m_2+...+m_N)}{\Gamma(m_1)\cdot\cdot\cdot \Gamma(m_N)}
 \eea
with $N=2$ and $m_1=\alpha$, $m_2=\beta$. Next we integrate over one
of the Feynman parameters, e.g. $x_2$, and subsequently redefine the
undetermined wave-number $\textbf{q} \rightarrow \textbf{q} +
x_1\textbf{k}$.

Now, in order to integrate over $\textbf{q}$, we build on the
following formula
 \be
 \int \frac{d^d\textbf{q}}{(2\pi)^d} \frac{1}{\(z \, \textbf{q}^2 + \Delta \)^n}
 = {z^{-d/2} \over (4\pi)^{d/2}} \, {\Gamma(n-d/2) \over \Gamma(n)}
 \, \Delta^{d/2-n} \, ,
 \ee
computed by means of dimensional regularization. Thus, for instance,
differentiating it with respect to $z$ a definite number of times
and setting $z=1$, one obtains
 \bea
 \int \frac{d^d\textbf{q}}{(2\pi)^d} \frac{\textbf{q}^2}{\(\textbf{q}^2 + \Delta \)^n}
 &=&{d/2 \over (4\pi)^{d/2}} \, {\Gamma(n-d/2-1) \over \Gamma(n)}
 \, \Delta^{d/2-n+1} \, ,
 \non
 \int \frac{d^d\textbf{q}}{(2\pi)^d}
 \frac{\(\textbf{q}^2\)^2}{\(\textbf{q}^2 + \Delta \)^n}
 &=&{d(d+2) \over 4 (4\pi)^{d/2}} \, {\Gamma(n-d/2-2) \over \Gamma(n)}
 \, \Delta^{d/2-n+2} \, .
 \eea
As a final step we integrate over $x_1$.

Another set of useful identities is related to the following
$d-$dimensional Fourier transform
 \bea
 \int \frac{d^d\textbf{k}}{(2\pi)^d} \frac{e^{i\textbf{k}\textbf{r}}}{\(\textbf{k}^2 \)^{\al}}
 ={1 \over (4\pi)^{d/2}} \, {\Gamma(d/2-\al) \over \Gamma(\al)}
 \, \( {\textbf{r}^2 \over 4} \)^{\al-d/2} \, .
 \eea
Differentiating it with respect to $\textbf{r}$ yields
 \bea
 \int \frac{d^d\textbf{k}}{(2\pi)^d} ~ \frac{\textbf{k}_i}{\(\textbf{k}^2 \)^{\al}} ~ e^{i\textbf{k}\textbf{r}}
 &=& i \, x_i \, { \Gamma(d/2-\al+1) \over 2(4\pi)^{d/2} \Gamma(\al)}
 \, \( {\textbf{r}^2 \over 4} \)^{\al-d/2-1} \, ,
 \non
 \int \frac{d^d\textbf{k}}{(2\pi)^d} ~ \frac{\textbf{k}_i\textbf{k}_j}{\(\textbf{k}^2 \)^{\al}} ~ e^{i\textbf{k}\textbf{r}}
 &=& {\Gamma(d/2-\al+1) \over (4\pi)^{d/2} \Gamma(\al)}
 \,  \(  {\delta_{ij} \over 2 } + (\al-d/2-1) {x_i x_j\over \textbf{r}^2} \) \( {\textbf{r}^2 \over 4} \)^{\al-d/2-1}
 \non
 \eea
 \bea
 \int \frac{d^d\textbf{k}}{(2\pi)^d} ~
 \frac{\textbf{k}_i\textbf{k}_j\textbf{k}_l}{\(\textbf{k}^2 \)^{\al}} ~
 e^{i\textbf{k}\textbf{r}}
 &=& {i\,\Gamma(d/2-\al+2) \over 16(4\pi)^{d/2} \Gamma(\al)}\( {\textbf{r}^2 \over 4} \)^{\al-d/2-3}
 \non
 && \times \, \[  \textbf{r}^2(\delta_{il}x_j+\delta_{jl}x_i+\delta_{ij}x_l) - (d-2\al+4) x_i x_j x_l\] \, ,
 \eea
 \begin{multline}
 \int \frac{d^d\textbf{k}}{(2\pi)^d} ~
 \frac{\textbf{k}_i\textbf{k}_j\textbf{k}_l\textbf{k}_m}{\(\textbf{k}^2 \)^{\al}} ~
 e^{i\textbf{k}\textbf{r}}
 = {\,\Gamma(d/2-\al+3) \over 32(4\pi)^{d/2} \Gamma(\al)}\( {\textbf{r}^2 \over 4} \)^{\al-d/2-4}[ \, (d-2\al+6)x_i x_j x_l x_m \,
 \\ - \textbf{r}^2
 (\delta_{im}x_jx_l+\delta_{jm}x_ix_l+\delta_{lm}x_ix_j+\delta_{il}x_mx_j+\delta_{jl}x_ix_m+\delta_{ij}x_lx_m) \\
 +{(\textbf{r}^2)^2 \over (d-2\al+4)} (\delta_{il}\delta_{jm}+\delta_{jl}\delta_{im}+\delta_{ij}\delta_{lm}) \, ] \, .
 \end{multline}

\end{document}